\documentclass[tight,twocolumn]{aastex62}
\usepackage{apjfonts}
\usepackage{float}
\usepackage[caption=false]{subfig}

\usepackage{graphics}
\usepackage{amssymb,amsmath}
\usepackage{color}

\newcommand{\host}{\hbox{WKK 6047}}

\def\lsim{\lower0.3em\hbox{$\,\buildrel <\over\sim\,$}}
\def\gsim{\lower0.3em\hbox{$\,\buildrel >\over\sim\,$}}

\newcommand{\msun}{\hbox{M$_{\odot}$}}

\newcommand{\halpha}{\hbox{H$\alpha$}}
\newcommand{\hbeta}{\hbox{H$\beta$}}
\newcommand{\hgamma}{\hbox{H$\gamma$}}
\newcommand{\hdelta}{\hbox{H$\delta$}}

\newcommand{\swift}{\textit{Swift}}

\begin{document}

\title{The Rise and Fall of ASASSN-18pg: Following a TDE from Early To Late Times}  
\shorttitle{The Rise and Fall of ASASSN-18pg} 
\shortauthors{Holoien, et al.}

\author[0000-0001-9206-3460]{Thomas~W.-S.~Holoien}
\altaffiliation{Carnegie Fellow}
\affiliation{The Observatories of the Carnegie Institution for Science, 813 Santa Barbara St., Pasadena, CA 91101, USA}

\author{Katie~Auchettl}
\affiliation{DARK, Niels Bohr Institute, University of Copenhagen, Lyngbyvej 2, 2100 Copenhagen, Denmark}
\affiliation{School of Physics, The University of Melbourne, Parkville, VIC 3010, Australia}
\affiliation{ARC Centre of Excellence for All Sky Astrophysics in 3 Dimensions (ASTRO 3D)}
\affiliation{Department of Astronomy and Astrophysics, University of California, Santa Cruz, CA 95064, USA}

\author[0000-0002-2471-8442]{Michael~A.~Tucker}
\altaffiliation{DOE Computational Science Graduate Fellow}
\affiliation{Institute for Astronomy, University of Hawai'i, 2680 Woodlawn Drive, Honolulu, HI 96822, USA}

\author[0000-0003-4631-1149]{Benjamin~J.~Shappee}
\affiliation{Institute for Astronomy, University of Hawai'i, 2680 Woodlawn Drive, Honolulu, HI 96822, USA}

\author{Shannon~G.~Patel}
\affiliation{The Observatories of the Carnegie Institution for Science, 813 Santa Barbara St., Pasadena, CA 91101, USA}

\author{James~C.~A.~Miller-Jones}
\affiliation{International Centre for Radio Astronomy Research---Curtin University, GPO Box U1987, Perth,
WA 6845, Australia}

\author{Brenna~Mockler}
\affiliation{Department of Astronomy and Astrophysics, University of California, Santa Cruz, CA 95064, USA}

\author{Dani\`el~N.~Groenewald}
\affiliation{South African Astronomical Observatory, PO Box 9, Observatory 7935, Cape Town, South Africa}
\affiliation{Southern African Large Telescope Foundation, PO Box 9, Observatory 7935, South Africa}

\author{Jonathan~S.~Brown}
\affiliation{Department of Astronomy and Astrophysics, University of California, Santa Cruz, CA 95064, USA}

\author{Christopher~S.~Kochanek}
\affiliation{Center for Cosmology and AstroParticle Physics (CCAPP), The Ohio State University, 191 W.\ Woodruff Ave., Columbus, OH 43210, USA}
\affiliation{Department of Astronomy, The Ohio State University, 140 West 18th Avenue, Columbus, OH 43210, USA}

\author{K.~Z.~Stanek}
\affiliation{Center for Cosmology and AstroParticle Physics (CCAPP), The Ohio State University, 191 W.\ Woodruff Ave., Columbus, OH 43210, USA}
\affiliation{Department of Astronomy, The Ohio State University, 140 West 18th Avenue, Columbus, OH 43210, USA}


\author{Ping~Chen}
\affiliation{Kavli Institute for Astronomy and Astrophysics, Peking University, Yi He Yuan Road 5, Hai Dian District, Beijing 100871, China}

\author{Subo~Dong}
\affiliation{Kavli Institute for Astronomy and Astrophysics, Peking University, Yi He Yuan Road 5, Hai Dian District, Beijing 100871, China}

\author{Jose~L.~Prieto}
\affiliation{N\'ucleo de Astronom\'ia de la Facultad de Ingenier\'ia y Ciencias, Universidad Diego Portales, Av. Ej\'ercito 441, Santiago, Chile}
\affiliation{Millennium Institute of Astrophysics, Santiago, Chile}

\author{Todd~A.~Thompson}
\affiliation{Center for Cosmology and AstroParticle Physics (CCAPP), The Ohio State University, 191 W.\ Woodruff Ave., Columbus, OH 43210, USA}
\affiliation{Department of Astronomy, The Ohio State University, 140 West 18th Avenue, Columbus, OH 43210, USA}

\author[0000-0002-1691-8217]{Rachael~L.~Beaton}
\altaffiliation{Hubble Fellow}
\altaffiliation{Carnegie-Princeton Fellow}
\affiliation{Department of Astrophysical Sciences, Princeton University, 4 Ivy Lane, Princeton, NJ~08544, USA}
\affiliation{The Observatories of the Carnegie Institution for Science, 813 Santa Barbara St., Pasadena, CA 91101, USA}

\author[0000-0002-7898-7664]{Thomas~Connor}
\affiliation{The Observatories of the Carnegie Institution for Science, 813 Santa Barbara St., Pasadena, CA 91101, USA}

\author[0000-0002-2478-6939]{Philip~S.~Cowperthwaite}
\affiliation{The Observatories of the Carnegie Institution for Science, 813 Santa Barbara St., Pasadena, CA 91101, USA}
\altaffiliation{Hubble Fellow}

\author{Linnea Dahmen}
\affiliation{Department of Physics \& Astronomy, Pomona College, 610 N College Ave, Claremont, CA 91711, USA}

\author[0000-0002-4235-7337]{K.~Decker~French}
\altaffiliation{Hubble Fellow}
\affiliation{The Observatories of the Carnegie Institution for Science, 813 Santa Barbara St., Pasadena, CA 91101, USA}

\author{Nidia~Morrell}
\affiliation{Las Campanas Observatory, Carnegie Observatories, Casilla 601, La Serena, Chile}

\author{David~A.~H.~Buckley}
\affiliation{South African Astronomical Observatory, PO Box 9, Observatory 7935, Cape Town, South Africa}

\author{Mariusz~Gromadzki}
\affiliation{Astronomical Observatory, University of Warsaw, Al. Ujazdowskie 4, 00-478 Warszawa, Poland}

\author{Rupak~Roy}
\affiliation{The Inter-University Centre for Astronomy and Astrophysics, Ganeshkhind, Pune - 411007, India}

\author{David~A.~Coulter}
\affiliation{Department of Astronomy and Astrophysics, University of California, Santa Cruz, CA 95064, USA}

\author{Georgios~Dimitriadis}
\affiliation{Department of Astronomy and Astrophysics, University of California, Santa Cruz, CA 95064, USA}

\author{Ryan~J.~Foley}
\affiliation{Department of Astronomy and Astrophysics, University of California, Santa Cruz, CA 95064, USA}

\author{Charles~D.~Kilpatrick}
\affiliation{Department of Astronomy and Astrophysics, University of California, Santa Cruz, CA 95064, USA}

\author{Anthony~L.~Piro}
\affiliation{The Observatories of the Carnegie Institution for Science, 813 Santa Barbara St., Pasadena, CA 91101, USA}

\author{C\'esar~Rojas-Bravo}
\affiliation{Department of Astronomy and Astrophysics, University of California, Santa Cruz, CA 95064, USA}

\author{Matthew~R.~Siebert}
\affiliation{Department of Astronomy and Astrophysics, University of California, Santa Cruz, CA 95064, USA}

\author{Sjoert~van~Velzen}
\affiliation{Center for Cosmology and Particle Physics, New York University, NY 10003}

\correspondingauthor{T.~W.-S.~Holoien}
\email{tholoien@carnegiescience.edu}

\date{\today}

\begin{abstract}
We present nearly 500 days of observations of the tidal disruption event ASASSN-18pg, spanning from 54 days before peak light to 441 days after peak light. Our dataset includes X-ray, UV, and optical photometry, optical spectroscopy, radio observations, and the first published spectropolarimetric observations of a TDE. ASASSN-18pg was discovered on 2018 July 11 by the All-Sky Automated Survey for Supernovae (ASAS-SN) at a distance of $d=78.6$ Mpc, and with a peak UV magnitude of $m\simeq14$ it is both one of the nearest and brightest TDEs discovered to-date. The photometric data allow us to track both the rise to peak and the long-term evolution of the TDE. ASASSN-18pg peaked at a luminosity of $L\simeq2.2\times10^{44}$ erg s$^{-1}$, and its late-time evolution is shallower than a flux $\propto t^{-5/3}$ power-law model, similar to what has been seen in other TDEs. ASASSN-18pg exhibited Balmer lines and spectroscopic features consistent with Bowen fluorescence prior to peak which remained detectable for roughly 225 days after peak. Analysis of the two-component \halpha{} profile indicates that, if they are the result of reprocessing of emission from the accretion disk, the different spectroscopic lines may be coming from regions between $\sim10$ and $\sim60$ light-days from the black hole. No X-ray emission is detected from the TDE and there is no evidence of a jet or strong outflow detected in the radio. Our spectropolarimetric observations give no strong evidence for significant asphericity in the emission region, with the emission region having an axis ratio of at least $\sim0.65$.
\end{abstract}
\keywords{accretion, accretion disks --- black hole physics --- galaxies: nuclei}


\section{Introduction}
\label{sec:intro}

When a star passes too close to a supermassive black hole (SMBH) and crosses its tidal radius, the tidal shear forces from the SMBH overwhelm the self-gravity of the star, resulting in a tidal disruption event (TDE). For a main-sequence star, roughly half of the stellar material remains bound to the SMBH, initially falling back to pericenter at a rate proportional to $\sim t^{-5/3}$. A fraction of this material is accreted onto the SMBH, resulting in a luminous, short-lived flare \citep[e.g.,][]{lacy82,rees88,evans89,phinney89}. 

Initial theoretical work predicted that the emission from the TDE flare would peak at soft X-ray energies and that the luminosity would evolve at a rate proportional to the $t^{-5/3}$ mass fallback rate. Recent studies of TDEs, however, have revealed that TDEs exhibit a wide range of observational properties \citep[e.g.,][]{velzen11,cenko12a,gezari12b,arcavi14, chornock14,holoien14b,gezari15,vinko15,holoien16a,holoien16b,brown16a,auchettl17,blagorodnova17,brown17a,gezari17,brown18,holoien18a,holoien19b,holoien19c,velzen19,leloudas19,velzen20}. We now know that the initial theoretical picture of TDE emission was too simplistic, as the emission depends on many factors, ranging from the disrupted star's physical properties \citep[e.g.,][]{macleod12,kochanek16}, the way the accretion stream evolves after disruption \citep[e.g.,][] {kochanek94,strubbe09,guillochon13,hayasaki13,hayasaki16,piran15,shiokawa15}, radiative transfer effects \citep[e.g.,][]{gaskell14,strubbe15,roth16,roth18}, and viewing angle \citep[e.g.,][]{dai18}. Despite the increasing number of known TDE flares, few have been observed in sufficient detail to differentiate between various theoretical predictions. In particular, very few TDEs have been discovered prior to peak light, making it difficult to study the early evolution of the stellar debris and the formation of the accretion disk.

Here we present the discovery and follow-up observations of ASASSN-18pg, a TDE flare discovered by the All-Sky Automated Survey for Supernovae \citep[ASAS-SN;][]{shappee14} on 2018 July 11 in the galaxy \host. We announced the discovery of the transient on 2018 July 15 on the Transient Name Server (TNS), where it was given the designation AT 2018dyb\footnote{\url{https://wis-tns.weizmann.ac.il/object/2018dyb}}, noting that the ASAS-SN position of the transient was consistent with the nucleus of the presumed host galaxy. We obtained an optical spectrum on 2018 July 17 \citep{asassn18pg_spec_atel} and found that the transient exhibited a strong blue continuum and several broad emission features, notably hydrogen Balmer and helium I and II lines, which are features consistent with a TDE \citep[e.g.,][]{arcavi14}.

After classifying ASASSN-18pg as a possible TDE, we requested and were awarded target-of-opportunity (TOO) observations from the \textit{Neil Gehrels Swift Gamma-ray Burst Mission} \citep[\swift;][]{gehrels04} UltraViolet and Optical Telescope \citep[UVOT;][]{roming05} and X-ray Telescope \citep[XRT;][]{burrows05} (Target ID: 10764). The {\swift} observations confirmed that the transient was UV-bright, but we did not detect any X-ray emission. Based on the spectra and UV-brightness of the source, we began an extended multi-wavelength campaign to monitor and characterize the emission of ASASSN-18pg. Due to the early detection and prompt announcement of discovery by the ASAS-SN team, we were able to begin follow-up data collection from {\swift} and various ground-based observatories well before the peak of the TDE's light curve, providing us with a rising light curve spanning from the $i$-band to the {\swift} UV filters and beginning 41 days prior to peak light. ASASSN-18pg thus provides us with one of the best opportunities to study the early emission from a TDE. We note that while early observations of ASASSN-18pg were the subject of a study by \citet{leloudas19}, their study was primarily focused on spectroscopic evolution of the TDE, while our dataset contains considerably more photometric data, and our treatment of the host galaxy (see Section~\ref{sec:archival}) provides for more robust host flux removal, allowing us to perform more extensive analyses. We also present spectropolarimetric observations of ASASSN-18pg obtained with the Southern African Large Telescope \citep[SALT;][]{buckley06}, the first such observations of a TDE.

In Section~\ref{sec:obs} we describe the pre-disruption data available for \host{} and fit its physical properties. We also discuss our follow-up observations of the transient and the available pre-discovery data available from ASAS-SN. In Section~\ref{sec:phot_anal} we analyze the photometric data, fit the light curves with a TDE emission model, model the blackbody evolution of ASASSN-18pg, and compare it to other TDEs. In Section~\ref{sec:spec_anal} we analyze the evolution of spectroscopic emission lines in ASASSN-18pg and discuss the results of spectropolarimetric observations of the transient taken near peak light. Finally, in Section~\ref{sec:disc} we summarize our findings and discuss the implications for future TDE studies.


\section{Observations}
\label{sec:obs}


\subsection{Archival Data and Host Fits}
\label{sec:archival}

Due to its southern declination, {\host} was not previously observed by optical surveys such as the Sloan Digital Sky Survey (SDSS) or Pan-STARRS. However, we were able to retrieve archival observations of the host in the $gri$ filters obtained with DECam mounted on the Blanco 4-m telescope at Cerro Tololo Inter-American Observatory in Chile in 2018 May as part of the ``Mapping Dust in 3D with DECam: A Galactic Plane Survey'' (Prop. ID 2018A-0251, PI D. Finkbeiner) from the NOAO Data Lab. We also obtained archival $JHK_S$ data from the Two Micron All-Sky Survey (2MASS) and in the $W1$ and $W2$ filters from the Wide-field Infrared Survey Explorer \citep[WISE;][]{wright10} AllWISE data release \citep{cutri13}. The host is not detected in archival data from, or was not previously observed by, the Galaxy Evolution Explorer (GALEX), Spitzer, Herschel, the Hubble Space Telescope (HST), the Chandra X-ray Observatory, the X-ray Multi-Mirror Mission (XMM-Newton), or the Very Large Array Faint Images of the Radio Sky at Twenty-cm (VLA FIRST) survey. 


\begin{figure*}
\begin{minipage}{\textwidth}
\centering
{\includegraphics[width=0.95\textwidth]{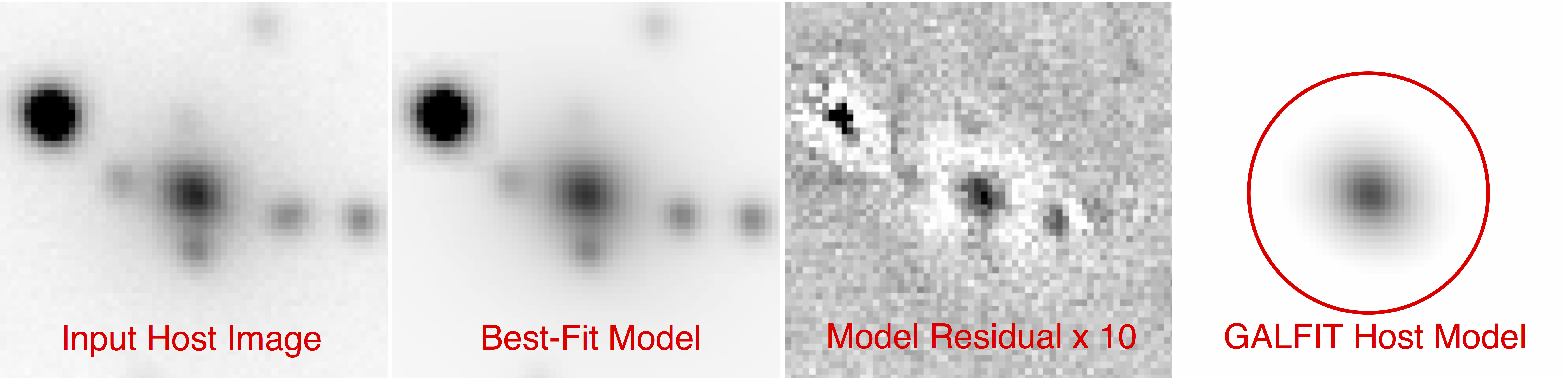}}
\caption{\emph{Left Panel:} The DECam $g$-band image of {\host} and the nearby stars. \emph{Center-left panel:} The best-fit star+galaxy model from GALFIT. \emph{Center-right panel:} The residuals of the model. The residuals have been increased by a factor of 10 to make them more visible. \emph{Right panel:} The GALFIT model of the host galaxy isolated from nearby stars. The red circle shows the aperture used to measure an aperture magnitude of the host, and has a radius equal to the $g$-band effective radius of the galaxy. All 4 images use the same scaling.}
\label{fig:host_ims}
\end{minipage}
\end{figure*}

The field of the host is heavily contaminated by Galactic sources, including 2 bright stars located off each end of the host galaxy's disk and 4 stars located in front of the host. While the 4 stars obstructing the host do not appear to contribute significant flux to the infrared (IR) data, the 2 nearby bright stars contaminate the 2MASS and AllWISE catalog magnitudes. To avoid contamination from these sources and obtain IR magnitudes of the host galaxy, we downloaded the $JHK_S$ 2MASS images and measured 5\farcs{0} aperture magnitudes in each filter. Unfortunately, we were unable to perform a similar analysis with the WISE data, as one of the bright stars was too bright to separate from the host galaxy. 

In the $gri$ DECam data, the stars in front of the host contribute a significant amount of flux, and we cannot measure the host flux directly. In order to obtain an estimate of the uncontaminated host flux in these filters, we used GALFIT \citep{peng02} to determine the flux of the host galaxy. Neighboring and obstructing stars were simultaneously included in the fit. SExtractor \citep{bertin96} was run on each band in order to measure object positions and magnitudes, which serve as initial guesses for GALFIT. A nearby, isolated, bright star was used as a PSF model. The sky mode was measured in each band following \citet{patel17} and used as a fixed estimate of the sky background. While the best-fitting Sersic index $n$, half-light radius $R_e$, and total magnitude are sensitive to the sky measurement (as well as other lingering artifacts in the imaging), the flux within $R<n\times R_e$ is fairly stable. We therefore use the best-fitting Sersic model for the TDE host galaxy to carry out aperture photometry using the effective radius of the host in each filter as the aperture radius, resulting in a robust measurement of the galaxy magnitudes. Aperture photometry was computed for each model image using the IRAF {\tt apphot} package, with the magnitudes being calibrated using multiple stars in the field of the host galaxy with known magnitudes in the AAVSO Photometric All-Sky Survey \citep[APASS;][]{henden15}. The input DECam $g$-band data, GALFIT galaxy+star model, model residuals, and isolated galaxy model are shown in Figure~\ref{fig:host_ims}, with the aperture used to measure the host magnitude shown in the right panel. We list the estimated $griJHK_S$ magnitudes in Table~\ref{tab:host_mags}.


\begin{deluxetable}{ccc}
\tabletypesize{\footnotesize}
\tablecaption{Archival Photometry of \host}
\tablehead{
\colhead{Filter} &
\colhead{Magnitude} &
\colhead{Magnitude Uncertainty} }
\startdata
$g$ & 16.58 & 0.11 \\
$r$ & 15.59 & 0.11 \\
$i$ & 15.21 & 0.11 \\ 
$J$ & 14.36 & 0.05 \\
$H$ & 14.05 & 0.04 \\
$K_S$ & 14.34 & 0.05
\enddata 
\tablecomments{Archival aperture magnitudes of {\host} measured from the GALFIT host model with aperture radius equal to the effective radius of the galaxy ($gri$) and from 2MASS data with 5\farcs{0} aperture radius ($JHK_S$). These magnitudes were used as the inputs for host-galaxy SED fitting.} 
\label{tab:host_mags} 
\end{deluxetable}

After measuring the $griJHK_S$ host magnitudes, we fit a spectral energy distribution (SED) to the host magnitudes using the publicly available Fitting and Assessment of Synthetic Templates \citep[\textsc{fast};~][]{kriek09} code. This fit assumed a \citet{cardelli89} extinction law with $R_V=3.1$ and a Galactic extinction of $A_V = 0.624$ mag \citep{schlafly11}. We adopted a Salpeter initial mass function, an exponentially declining star-formation history, and the \citet{bruzual03} stellar population models for the fit. Based on the \textsc{fast} fit, we find that the host has a stellar mass of $M_{\star}=1.7^{+0.1}_{-0.2} \times 10^{10}$ M$_{\odot}$, an age of $8.9^{+1.1}_{-1.8}$ Gyr, and an upper limit on the star formation rate of $\textrm{SFR}<2.6 \times 10^{-2}$ M$_{\odot}$~yr$^{-1}$. Scaling the stellar mass using the average stellar-mass-to-bulge-mass ratio from the hosts of ASASSN-14ae, ASASSN-14li, and ASASSN-15oi \citep{holoien14b,holoien16a,holoien16b}, as we have done with previous TDEs \citep[e.g.,][]{holoien19b}, gives an estimated bulge mass of $M_B\simeq10^{9.6}$~{\msun}. We then convert this to an estimated black hole mass of $M_{BH}=10^{7.0}$~{\msun} using the $M_B-M_{BH}$ relation from \citet{mcconnell13}. This is comparable to the masses of other TDE host galaxies \citep[e.g.,][]{holoien14b,holoien16a,holoien16b,brown17a,wevers17,mockler19} and our host properties in general are similar to those found by \citet{leloudas19}, with the exception of the SFR. However, \citet{leloudas19} based their host fit on small aperture optical and IR magnitudes, rather than the host flux fitting method we used, and they acknowledge that the SFR is not well-constrained in their analysis.

In order to obtain fluxes for the transient with the contributions from the host galaxy and nearby stars removed, we require measurements or estimates of the host and stellar fluxes in each filter contained in the 5\farcs{0} apertures used to measure transient magnitudes in our photometric follow-up campaign. For the $gri$ bands, we can measure a 5\farcs{0} host$+$star aperture magnitude directly from the archival DECam images to measure the total contaminating flux. For the {\swift} UVOT and $uBV$ data, however, we do not have any archival images from which to measure this flux. To approximate the host galaxy contribution in each filter, we used the \textsc{fast} SED fit of \host{} to derive synthetic 5\farcs{0} aperture magnitudes in each of our follow-up filters. We assume 10\% uncertainties on the host galaxy fluxes in each filter.

To estimate the contribution from the 4 stars contained in our aperture, we transformed the $gri$ PSF magnitudes from our GALFIT model using various transforms. For $B$- and $V$-band data, we used the \citet{lupton05} conversions to convert the $g$ magnitudes and $g-r$ colors into $B$ and $V$ magnitudes. For $u$-band, we used the $u-g$ colors of a large sample of stars in SDSS DR14 with similar $g-r$ and $r-i$ colors to each contaminating star to estimate the $u-g$ color of each of the 4 stars, and obtain a $u$-band magnitude from this. As there are no published transformations from the SDSS filters to {\swift} $U$-band, we assume the stars have the same fluxes and magnitudes in $U$ as they do in $u$. Finally, we ignore any contribution from the contaminating stars for the UVOT UV filters, as they do not appear to significantly contaminate the data in any epoch. The combined host$+$star 5\farcs{0} aperture magnitudes that we later subtracted from our follow-up data are shown for each filter in Table~\ref{tab:synth_mags}.


\begin{deluxetable}{ccc}
\tabletypesize{\footnotesize}
\tablecaption{5\farcs{0} Host$+$Star Aperture Magnitudes}
\tablehead{
\colhead{Filter} &
\colhead{Magnitude} &
\colhead{Magnitude Uncertainty} }
\startdata
$UVW2$ & 23.02 & 0.11 \\
$UVM2$ & 23.45 & 0.11 \\
$UVW1$ & 21.33 & 0.11 \\ 
$U_{UVOT}$ & 18.68 & 0.08 \\
$u$ & 18.57 & 0.08 \\
$B$ & 17.01 & 0.09 \\
$g$ & 16.39 & 0.09 \\
$V$ & 15.98 & 0.09 \\
$r$ & 15.43 & 0.09 \\
$i$ & 15.07 & 0.08
\enddata 
\tablecomments{5\farcs{0} aperture magnitudes of {\host} and the 4 contaminating stars contained in the aperture synthesized for the {\swift} UV$+U$ and $uBV$ filters as described in Section~\ref{sec:archival} and measured directly for the $gri$ filters. All magnitudes are in the AB system.} 
\label{tab:synth_mags} 
\end{deluxetable}

\subsection{ASAS-SN light curve}
\label{sec:ASASSN_LC}

ASAS-SN uses units of four 14-cm telescopes on a common mount to monitor the full visible sky on a rapid cadence to find bright, nearby transients \citep{shappee14,kochanek17}. ASAS-SN currently is composed of five units hosted by the Las Cumbres Observatory global telescope network \citep{brown13} in Hawaii, Chile, Texas, and South Africa. New ASAS-SN images are processed using a fully automatic pipeline that incorporates the ISIS image subtraction package \citep{alard98, alard00}. To obtain photometry of ASASSN-18pg uncontaminated by the host and nearby stars, we constructed a reference image of the host galaxy and surrounding sky for each ASAS-SN unit that could observe it. ASASSN-18pg was discovered when the two original ASAS-SN units were still using $V$ filters and the new $g$-band telescopes were still building images for references rather than performing normal survey operations. Because of this, we have several years' worth of data of the field in $V$-band, but no images in $g$-band more than a few weeks prior to discovery, when it is likely the images would contain some transient flux. To construct the $V$-band reference image, we used only data obtained prior to 2018 May 01, and for the $g$-band reference image, we used only data obtained after 2019 April 01, when the transient flux was no longer apparent in our data. 

\begin{figure*}
\begin{minipage}{\textwidth}
\centering
{\includegraphics[width=0.95\textwidth]{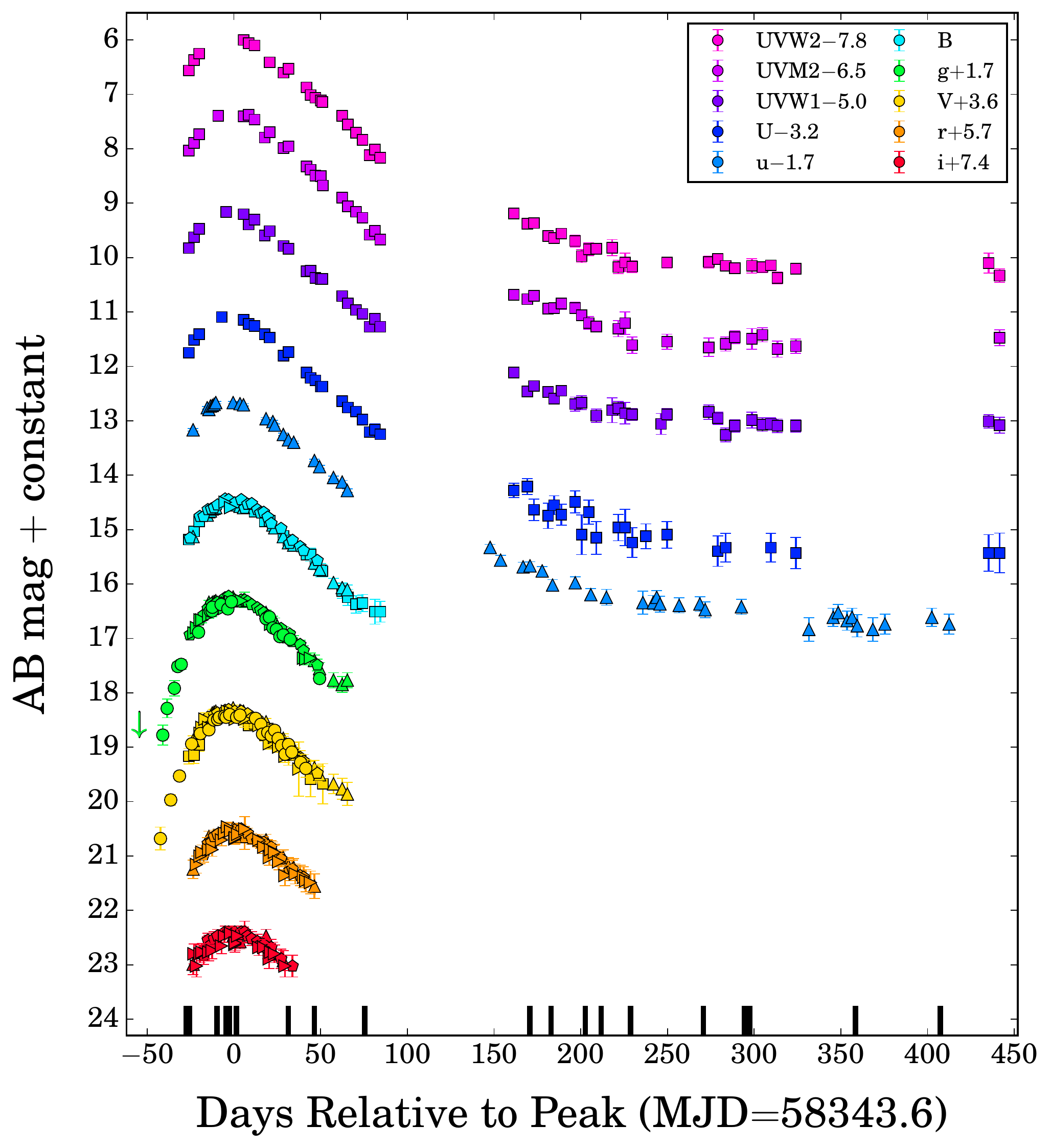}}
\caption{Host-subtracted UV and optical light curves of ASASSN-18pg spanning from 54 days prior to peak brightness (MJD$=58343.6$, measured from the $g$-band light curve; see Section~\ref{sec:params}) to 441 days after peak. ASAS-SN ($gV$) data are shown as circles; {\swift} UVOT data are shown as squares; and Swope ($uBgVri$), Las Cumbres Observatory 0.4-m ($BVgri$), and Las Cumbres Observatory 1-m ($BVgri$) data are shown as triangles, right-facing triangles, and pentagons, respectively. $3\sigma$ upper limits are shown with downward arrows. Early ASAS-SN data have error bars in time to denote the date range of observations that have been combined to obtain a single measurement, though in some cases these error bars may be smaller than the points. {\swift} $B$- and $V$-band data were converted to Johnson $B$ and $V$ magnitudes to enable direct comparison with the ground-based data. Black bars along the bottom of the Figure show epochs of spectroscopic follow-up. All data have been corrected for Galactic extinction and are presented in the AB system.}
\label{fig:lc}
\end{minipage}
\end{figure*}

We then used these references to subtract the background and host emission from all science images. We performed aperture photometry on each host-template subtracted image using the {\sc Iraf} {\tt apphot} package, and calibrated the magnitudes to several stars in the vicinity of the transient with known magnitudes in the AAVSO Photometric All-Sky Survey \citep[APASS;][]{henden15}. For some pre-discovery epochs, when ASASSN-18pg was still very faint, we stacked several science images to improve the signal-to-noise of our detections. We present the ASAS-SN photometry (detections and $3\sigma$ limits) in Table~\ref{tab:phot} and include them in Figure~\ref{fig:lc}. We use error bars on the X-axis to denote the date ranges of epochs that were combined to obtain higher signal-to-noise measurements.


\begin{deluxetable}{cccc}
\tabletypesize{\footnotesize}
\tablecaption{Host-Subtracted Photometry of ASASSN-18pg}
\tablehead{
\colhead{MJD} &
\colhead{Filter} &
\colhead{Magnitude} &
\colhead{Telescope} }
\startdata
58320.07 & $i$ & $15.59\pm0.19$ & Swope \\
58320.58 & $i$ & $15.40\pm0.18$ & LCOGT\_04m \\  
58322.01 & $i$ & $15.62\pm0.20$ & LCOGT\_04m \\ 
... & & & \\
58653.23 & $W2$ & $17.94\pm0.09$ & \swift \\  
58593.33 & $W2$ & $18.17\pm0.11$ & \swift \\
58617.42 & $W2$ & $18.01\pm0.09$ & \swift 
\enddata 
\tablecomments{Host-subtracted magnitudes and $3\sigma$ upper limits for all photometric follow-up data. The Telescope column indicates the source of the data for each epoch: ``ASAS-SN'' is used for ASAS-SN survey data, ``Swope'' is used for data from the 1-m Swope telescope at Las Campanas Observatory, ``LCOGT\_04m'' and ``LCOGT\_1m'' are used for data from the Las Cumbres Observatory 0.4-m and 1-m telescopes, respectively, and ``\swift'' is used for {\swift} UVOT data. All measurements have been corrected for Galactic extinction and are presented in the AB system. Only a portion of this Table is shown here, for guidance regarding its form and content; the entire table is published in machine-readable format in the online journal.}
\label{tab:phot} 
\end{deluxetable}


\subsection{Swift Observations}
\label{sec:swift}

Our initial {\swift} follow-up campaign included 25 epochs of TOO observations between 2018 July 18 and 2018 November 06, when ASASSN-18pg became Sun-constrained. After it re-emerged from behind the Sun, we obtained an additional 28 epochs of observations between 2019 January 22 and 2019 October 29, when it became Sun-constrained again. UVOT observations were obtained in the $V$ (5468 \AA), $B$ (4392 \AA), $U$ (3465 \AA), $UVW1$ (2600 \AA), $UVM2$ (2246 \AA), and $UVW2$ (1928 \AA) filters \citep{poole08} in most epochs, with some epochs having fewer filters, depending on scheduling. Since each epoch contains 2 observations in each filter, we first combined the two images in every filter using the HEAsoft software task {\tt uvotimsum}, then extracted counts from the combined images in a 5\farcs{0} radius region using the software task {\tt uvotsource}, using a sky region of $\sim$~40\farcs{0} radius to estimate and subtract the sky background. We calculated magnitudes and fluxes from the UVOT count rates based on the most recent UVOT calibration \citep{poole08,breeveld10}. 

We assumed a \citet{cardelli89} extinction law to correct the UVOT transient, host, and star magnitudes for Galactic extinction, then subtracted the 5\farcs{0} host$+$star fluxes from each observation to isolate the transient flux in each epoch. In order to directly compare the {\swift} $B$- and $V$-band data to our ground-based observations, we converted the UVOT $B$ and $V$ magnitudes to Johnson $B$ and $V$ magnitudes using publicly available color corrections\footnote{\url{https://heasarc.gsfc.nasa.gov/docs/heasarc/caldb/swift/docs/uvot/uvot_caldb_coltrans_02b.pdf}}. The {\swift} UVOT photometry are shown in Figure~\ref{fig:lc} and presented in Table~\ref{tab:phot}.

ASASSN-18pg was also observed using the photon counting mode of {\swift}'s X-Ray Telescope (XRT). All observations were reprocessed using the {\swift} analysis tool {\tt xrtpipeline} version 0.13.2, using the standard filters and screening suggested by the {\swift} data reduction guide\footnote{\url{https://swift.gsfc.nasa.gov/analysis/xrt_swguide_v1_2.pdf}} and the most up to date CALDB.  To place constraints on the presence of X-ray emission arising from ASASSN-18pg, we used a source region centered on the position of ASASSN-18pg with a radius of 30 arcseconds, and a source free background region centered at ($\alpha$, $\delta$)=(16:18:35.3,$-$61:00:48.4). Similar to \citet{leloudas19}, we find no significant X-ray emission from the source during its evolution. In order to place the strongest constraints on the X-ray emission arising from ASASSN-18pg, we merged all 54 observations of ASASSN-18pg using {\tt xselect} version 2.4g. We derived a 3$\sigma$ upper limit on the count rate of 0.001 counts/sec for the 0.3-10.0 keV energy range. Assuming an absorbed blackbody model with a temperature of 0.05 keV similar to that of other X-ray bright TDEs \citep[e.g., ASASSN-14li, ASASSN-15oi][]{brown16a, holoien18a} at the redshift of the host galaxy and a Galactic column density of $1.77\times10^{21}$ cm$^{-2}$ \citep{hi4pi16}, we obtain an absorbed flux of $2.6\times10^{-14}$ erg cm$^{-2}$ s$^{-1}$, which corresponds to an upper limit on the X-ray luminosity of $L_X\sim2\times10^{40}$ erg s$^{-1}$.

We do detect weak ($\sim$2$\sigma$ above background) X-ray emission observed during observations ObsID:00010764017 and ObsID:00010764027. Here we find a background subtracted count rate in the 0.3-10.0 keV range that has been corrected for encircled energy fraction of 0.004$\pm$0.002 count~s$^{-1}$ and 0.003$\pm$0.001 count~s$^{-1}$ for ObsID:00010764017 and ObsID:00010764027, respectively. Assuming the same absorbed blackbody model that we used to derive the 3$\sigma$ upperlimit from the merged observations, we get an absorbed luminosity of $(9\pm5)\times10^{40}$ erg s$^{-1}$ and $(7\pm3)\times10^{40}$ erg s$^{-1}$, respectively. This is $\sim4$ orders of magnitude less than the bolometric luminosity detected at peak. Assuming the BH mass derived in Section \ref{sec:archival}, this suggests that the source is emitting X-rays at only $\sim0.01\%$ of Eddington, consistent with what has been found from other X-ray emitting TDEs \citep[e.g.,]{mockler19,wevers19}. 

\subsection{Other Photometric Observations}
\label{sec:other_phot_obs}

We also obtained $uBVgri$ observations from the Swope 1-m telescope at Las Campanas Observatory and $BVgri$ observations from the Las Cumbres Observatory 0.4-m and 1-m telescopes located in Cerro Tololo, Chile; Siding Spring, Australia; and Sutherland, South Africa \citep{brown13}. We measured 5\farcs{0} aperture magnitudes in these data using the IRAF {\tt apphot} package, using a 13\farcs{0}$-$19\farcs{0} annulus to estimate and subtract background counts while avoiding the nearby contaminating stars. We used several stars in the field with magnitudes available in the APASS DR 10 catalog to calibrate the $BVgri$ data. For each comparison star, we estimated a $u$ magnitude by first calculating the average $u-g$ color of a large sample of SDSS DR14 stars with similar $g-r$ colors to the star in question, then assuming this $u-g$ color to estimate a $u$ magnitude using the APASS $g$-band magnitude. These $u$ magnitudes were then used to calibrate the $u$-band data.

As with the UVOT observations, we corrected all ground-based aperture magnitudes for Galactic extinction and subtracted the flux of the host galaxy and contaminating stars. The host-subtracted ground-based photometry are presented in Table~\ref{tab:phot} and shown in Figure~\ref{fig:lc}.

\subsection{Spectroscopic Observations}
\label{sec:spec_obs}

We began spectroscopic follow-up observations of ASASSN-18pg following its classification as a possible TDE and continued to monitor it regularly through 2019 September. Our follow-up spectra were obtained with the Robert Stobie Spectrograph \citep[RSS;][]{burgh03} on the 10-m SALT, the Gemini Multi-Object Spectrograph \citep[GMOS;][]{hook04,gimeno16} on the 8.4-m Gemini South telescope, the Inamori-Magellan Areal Camera and Spectrograph \citep[IMACS;][]{dressler11} on the 6.5-m Magellan-Baade telescope, LDSS-3 on the 6.5-m Magellan Clay telescope, the Goodman Spectrograph \citep{clemens04} on the Southern Astrophysical Research (SOAR) 4.1-m telescope, and the Wide Field Reimaging CCD Camera (WFCCD) on the du Pont 100-inch telescope. Our observations span from 26 days prior to peak light through 272 days after and include several spectra taken near or before peak light. 


\begin{figure*}
\begin{minipage}{\textwidth}
\centering
{\includegraphics[width=0.95\textwidth]{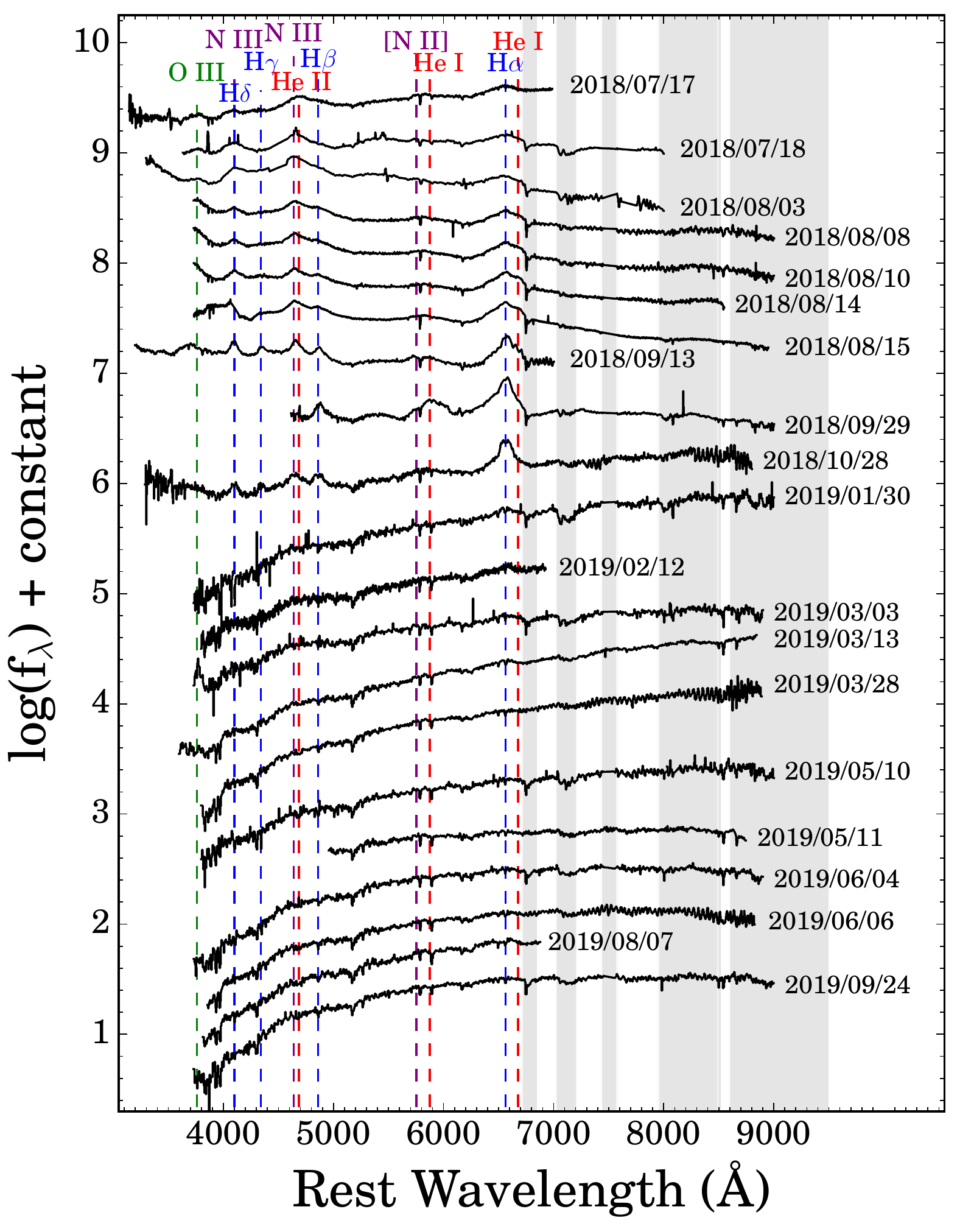}}
\caption{Spectroscopic evolution of ASASSN-18pg spanning from 26 days prior to peak (2018 August 12) through 407 days after peak. As described in Section~\ref{sec:spec_obs}, the spectra have been flux-calibrated using our photometric observations. The date each spectrum was taken is shown to the right of each spectrum, and prominent hydrogen, helium, nitrogen, and oxygen features identified by \citet{leloudas19} are indicated with blue, red, purple, and green dashed lines, respectively. Gray-shaded regions indicate telluric bands.}
\label{fig:spec_evol}
\end{minipage}
\end{figure*}

We reduced and calibrated the majority of our spectra using standard \textsc{Iraf} procedures, including bias subtraction, flat-fielding, 1-D spectrum extraction, and wavelength calibration with an arc lamp taken immediately before or after the science spectra. Most of our observations were then flux calibrated using spectrophotometric standard star spectra obtained on the same night as the science spectra. Spectra obtained with SOAR were flux calibrated using a custom IDL pipeline. Spectra obtained with SALT were reduced in part using the {\tt PySALT} software package \citep{crawford10}. Absolute flux calibration with SALT is difficult because of the telescope design, which has a moving, field-dependent and under-filled entrance pupil. Observations of spectrophotometric flux standards can, at best, only provide relative flux calibration \citep[see, e.g.,][]{buckley18}, which mostly accounts for the low frequency telescope and instrument sensitivity changes as a function of wavelength. We present the details of the spectra in our dataset in Table~\ref{tab:spec_details}.

We also used our photometric dataset to further flux calibrate our spectra. As our spectra were observed through slits of roughly 1\farcs{0} width, we measured magnitudes from our Swope and Las Cumbres Observatory data using a 1\farcs{5} aperture to obtain magnitudes with a similar amount of host contamination as would be present in our spectra. For all photometric filters that were completely contained in the wavelength range covered by a given spectrum and for which we could either interpolate the small aperture light curves or extrapolate them by 1 hour or less, we extracted synthetic photometric magnitudes from the spectrum. We then fit a line to the difference between the observed fluxes and the synthetic fluxes as a function of the central filter wavelength and scaled the spectra by the photometric fits. Finally, we corrected the spectra for Galactic extinction using a Milky Way extinction curve, assuming $R_V=3.1$ and $A_V=0.624$ \citep{schlafly11}.

Our final calibrated spectra of ASASSN-18pg are shown in Figure~\ref{fig:spec_evol}. We also mark prominent telluric bands in the Figure and have masked the telluric feature from 7550\AA$-$7720\AA\ and chip gaps (where present). Unlike what was seen in PS18kh \citep{holoien19b} and ASASSN-19bt \citep{holoien19c}, where the broad lines did not form until the TDEs were at or near peak light, ASASSN-18pg exhibited broad lines in all spectra taken prior to peak. This may indicate that the lines are being generated through different physical processes in ASASSN-18pg, and we further analyze the line emission in Section~\ref{sec:spec_anal}.

Our second SALT spectrum, obtained on 2018 August 03, was a low resolution (PG0300 grating) spectropolarimetric observation \citep{nordsieck03} obtained under clear conditions with an average seeing of $\sim$2\farcs{2}. Four 900s exposures were obtained at four half waveplate positions (0, 45, 22.5 and 67.5 degrees). The data reductions were carried out using the beta version of the {\tt polSALT} software\footnote{\url{https://github.com/saltastro/polsalt}}. The software perform basic image reductions on the raw SALT data, after which the data is then wavelength calibrated. The Stoke Q and U parameters, the magnitude of the linear polarization, p, and the position angle of the E-vector, $\theta$ are then determined. This is the first reported spectropolarimetric observation of a TDE, and we discuss the results further in Section~\ref{sec:specpol}.

\subsection{Radio Observations}
\label{sec:radio_obs}

We observed ASASSN-18pg using the Australia Telescope Compact Array (ATCA) in the 15mm band, using the Compact Array Broadband Backend \citep[CABB;][]{wilson11} to provide$2\times2048$\,MHz of bandwidth, centred at 16.7 and 21.2\,GHz.  Our initial observation was made on 2018 July 20 (08:18--13:29 UT), with the array in its compact H75 configuration, with the inner five antennas all within 90\,m, and the sixth antenna separated by 6\,km.

We used the standard flux density calibrator PKS B1934$-$638 to calibrate the bandpass and set the amplitude scale, and the nearby calibrator 1613$-$586 to solve for the complex antenna gains as a function of time. We reduced the data using standard procedures within the Common Astronomy Software Application \citep[CASA;][]{mcmullin07}. We imaged the data using Briggs weighting with a robustness parameter of 1, as a compromise between sensitivity and resolution. We reached an image noise level of 12\,$\mu$Jy\,beam$^{-1}$ by stacking both frequency bands. While the source position was coincident with a 50\,$\mu$Jy\,beam$^{-1}$ peak in the image, it was close enough to a brighter (0.7 mJy) nearby source at co-ordinates ($\alpha$,$\delta$)$=$(16:10:54.52, $-$60:56:04.8) that it could potentially be attributed to sidelobe confusion in this compact configuration, especially given its marginal ($<5\sigma$) significance.

To verify whether or not this marginal detection was real, we made a second ATCA observation on 2018 August 6 (12:20--17:18 UT), with the array in a more extended 1.5\,km configuration, providing significantly improved resolution to distinguish the target from the nearby confusing source.  We used the same observational setup and data analysis procedures, and did not detect a source at the target position down to a $3\sigma$ upper limit of 43\,$\mu$Jy\,beam$^{-1}$.  We therefore conclude that ASASSN-18pg was not detected in the radio.


\section{Photometric Analysis}
\label{sec:phot_anal}

\subsection{Position, Redshift, and $t_{Peak}$ Measurements}
\label{sec:params}

In order to measure the position of ASASSN-18pg, we first generated an image of the TDE by subtracting a $g$-band image from the Las Cumbres Observatory 1-m telescopes taken in 2019 July from a similar $g$-band image taken near peak. Using the \textsc{Iraf} task \texttt{imcentroid} we then measured a centroid position of the TDE flux in the subtracted image as well as the centroid position of the host galaxy nucleus in the archival $g$-band DECam image. The resulting position of ASASSN-18pg is ($\alpha$,$\delta$)$=$(16:10:58.89,$-$60:55:24.18), which is offset by 0\farcs{20} from the position of the host measured in the archival image. This offset is likely dominated by systematic offset in the astrometry between the two images. To account for this we also measured the centroid positions of several stars in both the pre-subtracted, peak $g$-band image and the archival host image and calculated an average offset for the positions of these comparison stars of 0\farcs{24}, with the stars being offset in various directions. Thus, the TDE is offset by $0\farcs{20}\pm0\farcs{24}$ from its host, corresponding to a physical offset of $75.9\pm91.1$~pc. 

The redshift of WKK 6047 was reported by \citet{woudt08} as $z=0.017392$. We also measured the redshift of the TDE using the \ion{Ca}{2} H \& K absorption features that are visible in the 2018 August 15 IMACS spectrum, finding $z=0.018$. As this is consistent with the \citet{woudt08} measurement, we adopt the archival $z=0.017392$, corresponding to a luminosity distance of $d=78.6$~Mpc, throughout the manuscript.


\begin{figure*}
\begin{minipage}{\textwidth}
\centering
{\includegraphics[width=0.95\textwidth]{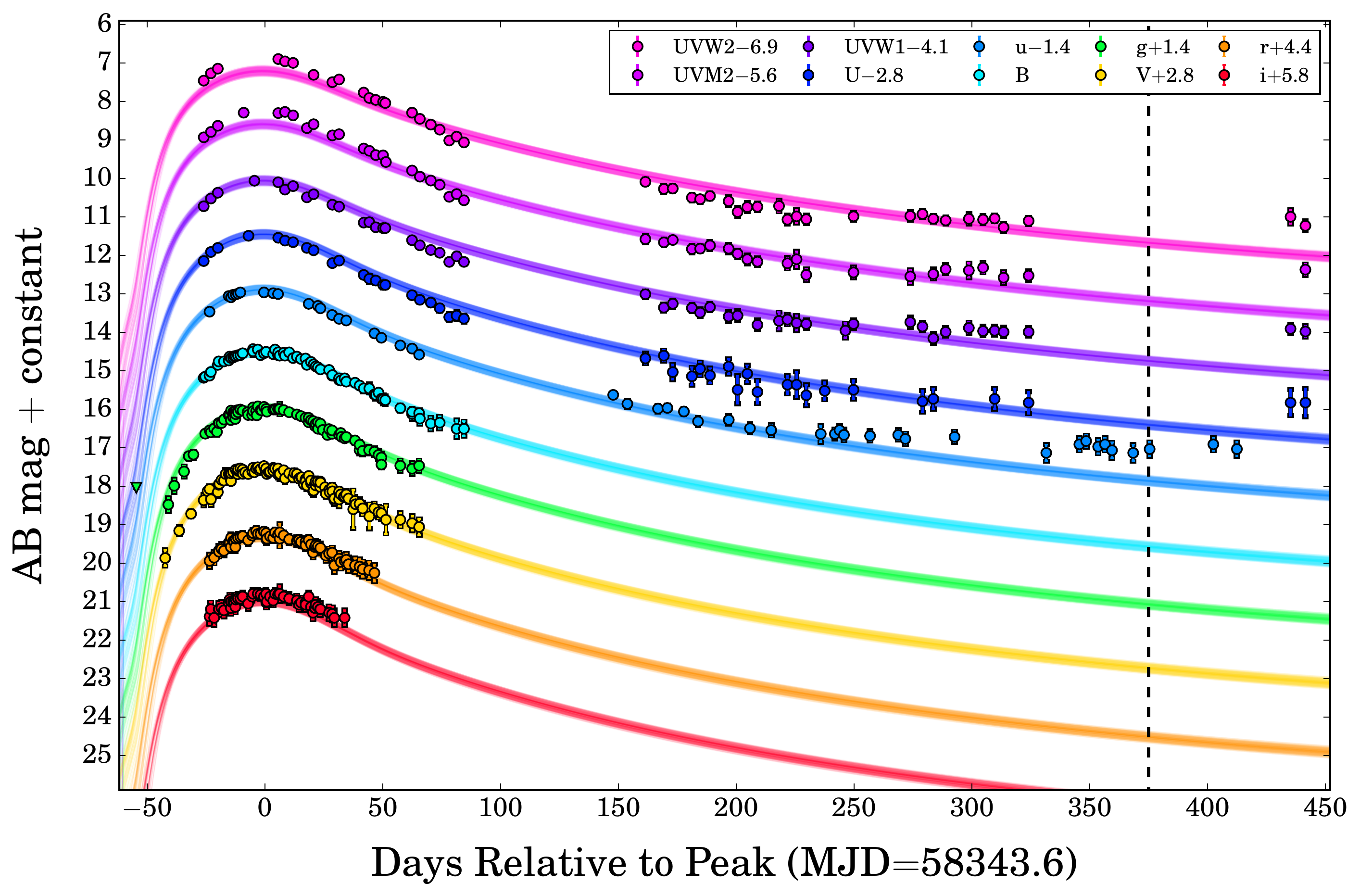}}
\caption{Multi-band light curve fits from \texttt{MOSFiT} with the host-subtracted light curves overplotted. The fits shown represent the $1-99$\% range of fits for each filter. Detections are shown as circles and $3\sigma$ upper limits are shown with downward triangles, and colors match those of Figure~\ref{fig:lc}. Data to the right of the dashed line were not included when performing the fit, as described in the text.}
\label{fig:mosfit_lc}
\end{minipage}
\end{figure*}

To estimate the time of peak light, we used the $g$-band light curve, which has the best sampling across the peak. We fit a parabola to the host-subtracted data from ASAS-SN and other ground-based telescopes taken between MJD$=$58320 and MJD$=$58370, as the declining light curve is flatter than the rising light curve, making a parabolic fit to the entire light curve impossible. To estimate the uncertainty on the peak time, we generated 10000 $g$-band light curves for our specified date range with each magnitude perturbed by its uncertainty, assuming Gaussian errors. We then used a parabolic fit to fit each of these 10000 light curves and calculated the 68\% confidence interval from each of these realizations. Based on this, we find $t_{g,peak}=58343.6\pm0.3$ and $m_{g,peak}=14.6$. Using the same procedure to calculate the peak times for each of our photometric filters, we find there is some evidence that the redder filters peaked later than the bluer filters, with $t_{UVW2,peak}=58340.8\pm0.4$ and $t_{i,peak}=58345.2\pm1.3$, similar to other TDEs \citep[e.g.,][]{holoien18a,holoien19b}. As the $g$-band light curve is the best-sampled (due to the ASAS-SN survey data in addition to our high cadence follow-up data), we adopt the $g$-band peak of $t_{g,peak}=58343.6$, corresponding to 2018 August 13.6, throughout our analysis.

\subsection{MOSFiT Light Curve Analysis}
\label{sec:lc_anal}

In order to extract physical parameters of ASASSN-18pg from our photometric dataset, we fit the multi-band host-subtracted light curves of the TDE using the Modular Open-Source Fitter for Transients \citep[\texttt{MOSFiT};][]{guillochon17}. \texttt{MOSFiT} generates bolometric light curves of transients using models that contain several physical parameters, uses these bolometric light curves to generate single-filter light curves, and fits these to the observed data. It then uses one of various sampling methods to find the combination of parameters that yield the highest likelihood match for a given model. We used the built-in TDE model to fit the light curves of ASASSN-18pg, and due to the large number of photometric filters and observations in our dataset, we ran \texttt{MOSFiT} in nested sampling mode for our fits. More details on \texttt{MOSFiT} and specifics on its TDE model can be found in \citet{guillochon17} and \citet{mockler19}.

While the \texttt{MOSFiT} TDE model lacks some physical parameters, such as an accretion disk module that can explain X-ray emission, it is the only tool available for generalized fitting of TDE emission, and works well for modeling cases such as ASASSN-18pg, where the light curves evolve smoothly and there is no X-ray emission. The \texttt{MOSFiT} multi-band fits to the ASASSN-18pg light curves are shown in Figure~\ref{fig:mosfit_lc} with our data overplotted. Our extremely well sampled light curves of ASASSN-18pg provide an excellent input dataset, and the \texttt{MOSFiT} fits match both the early and late-time data fairly well, though the fits do underpredict the emission in our latest epochs of observation. Comparing to the fits of several previous TDE discoveries in \citet{mockler19}, the rise of ASASSN-18pg is much better constrained than the majority of the TDEs in their sample, as we have significantly more data prior to and around peak light.

When fitting the \texttt{MOSFiT} model we only included observations obtained up to 375 days after peak. We found that when the latest few observations were included in the fits, the late-time data were better fit, but that the rising and peak parts of the light curve were fit significantly worse. Because the rise is so well-constrained by our data, we expect that the most likely explanation for \texttt{MOSFiT} being unable to fit both the early- and late-time data is that either our host flux subtraction method is slightly underpredicting the host emission, resulting in some host contamination that becomes more apparent as the transient emission fades, or that the \texttt{MOSFiT} TDE model does not incorporate the physical components needed to fit both the early- and late-time data simultaneously. Recent studies of TDEs \citep[e.g.,][]{brown17a,holoien18a,velzen19} have shown that the UV and bluer bands often flatten at later times, which has been attributed to a transition from fallback-dominated to disk-dominated emission. The \texttt{MOSFiT} TDE model was built to predict TDE emission when the bolometric luminosity closely follows the fallback rate, which is likely why it has trouble fitting the data at late times, when this is not necessarily the case. Therefore, we prioritized fitting the early-time data well over the late-time data.


\begin{deluxetable}{ccc}
\tabletypesize{\footnotesize}
\tablecaption{\texttt{MOSFiT} Model Parameter Fits}
\tablehead{
\colhead{Quantity} &
\colhead{Value} &
\colhead{Units} }
\startdata
$\log{R_{\textrm{ph0}}}$ & $1.00^{+0.48}_{-0.47}$ & --- \\
$\log{T_{\textrm{viscous}}}$ & $-0.69^{+0.89}_{-1.31}$ & days \\
$b$ (scaled $\beta$) & $1.02^{+0.36}_{-0.36}$ & --- \\
$\log{M_{BH}}$ & $7.18^{+0.23}_{-0.23}$ & \msun \\
$\log{\epsilon}$ (efficiency) & $-0.89^{+0.74}_{-0.74}$ & --- \\
$l$ (photosphere exponent) & $1.80^{+0.24}_{-0.23}$ & --- \\
$\log{n_{\textrm{H,host}}}$ & $20.74^{+0.02}_{-0.03}$ & cm$^{-2}$ \\
$M_\star$ & $0.10^{+0.36}_{-0.08}$ & \msun \\
$t_\textrm{exp}$ & $-21.38^{+16.40}_{-16.59}$ & days \\
$\log{\sigma}$ & $-0.89^{+0.02}_{-0.02}$ & --- \\
\enddata 
\tablecomments{Best-fit TDE model parameters from \texttt{MOSFiT} and $1-99$\% range on the uncertainties. Units are listed where appropriate. The uncertainties shown include the systematic uncertainties from Table 3 of \citet{mockler19}.}
\label{tab:mosfit_params} 
\end{deluxetable}

Table~\ref{tab:mosfit_params} shows the median values and $1-99$\% range for all the parameters of the \texttt{MOSFiT} TDE model. The model parameters are in general very well constrained, with the results suggesting that the star was almost certainly completely disrupted in the encounter. We note that the values reported in Table~\ref{tab:mosfit_params} include systematic uncertainties (see Table 3 of \citealp{mockler19}), and that in general the systematic uncertainties on the model parameters are much larger than the uncertainties from the fit.

After accounting for systematic uncertainties, the black hole mass is $M_{BH}=1.5^{+1.0}_{-0.6}\times10^7$~\msun, consistent with our estimate based on the stellar luminosity of the host in Section~\ref{sec:archival}. The mass of the disrupted star is $M_\star=0.10^{+0.36}_{-0.08}$~\msun, which is low but consistent with that of several other TDEs in \citet{mockler19}. This is of interest, as TDEs should occur more frequently with stars of $M\lesssim0.3$~{\msun} \citep{kochanek16}. 

To test the robustness of this fit, we also performed fits with the same data while adjusting the maximum photosphere size and the Eddington limit. Altering these parameters did not significantly affect the black hole mass, but did result in some changes to the photosphere parameters, an increase in the stellar mass, and a decrease in the efficiency. The systematic errors from \texttt{MOSFiT} are thus likely the primary source of uncertainty for these parameters of the model.

\citet{leloudas19} performed a similar fit with \texttt{MOSFiT} using only the early-time {\swift} UV data and found best-fit values of $M_{BH}=4^{+5}_{-2}\times10^6$~\msun and $M_\star=0.7^{+4.0}_{-0.6}$~\msun, marginally consistent with our results, although our results are significantly better constrained. We performed fits using both the same epochs of UV data used by \citet{leloudas19} and our full UV dataset without any optical data, finding in the former case that the mass was $M_{BH}=7.8^{+8.8}_{-4.1}\times10^6$~\msun and in the latter that the mass was $M_{BH}=1.1^{+1.1}_{-0.5}\times10^7$~\msun. While the black hole mass from our fit with the same UV data as \citet{leloudas19} is consistent with their black hole mass, it is substantially higher. In a private communication with G. Leloudas, we discovered a $0.2-0.3$ mag difference (a difference of roughly $\sim10$\%) in the Galactic extinction applied to correct the UV filters, with our calculated extinction values resulting in brighter magnitudes. We note that this is likely the source of the bulk of the discrepancy between our fits using the same epochs of {\swift} data.

The addition of the high-cadence optical data provides useful constraints on the rise time. This lowers the uncertainties on several physical quantities associated with the rising part of the light curve, in particular the star and black hole masses. This highlights the need for both UV and optical monitoring prior to peak light to properly constrain these parameters with tools like \texttt{MOSFiT}. 

\subsection{SED Analysis}
\label{sec:sed_anal}

As we have done with previous TDEs \citep[e.g.,][]{holoien19b,holoien19c}, we modeled the UV and optical SED of ASASSN-18pg as a blackbody for epochs where \swift{} data were available. We fit the SED using a flat temperature prior of $10000$~K~$\leq T \leq55000$~K and used Markov Chain Monte Carlo methods to fit the blackbody SED to the data in each epoch. We then estimated the bolometric luminosity, temperature, and radius of ASASSN-18pg in each epoch from the SED fits.


\begin{figure*}
\begin{minipage}{\textwidth}
\centering
\subfloat{{\includegraphics[width=0.48\textwidth]{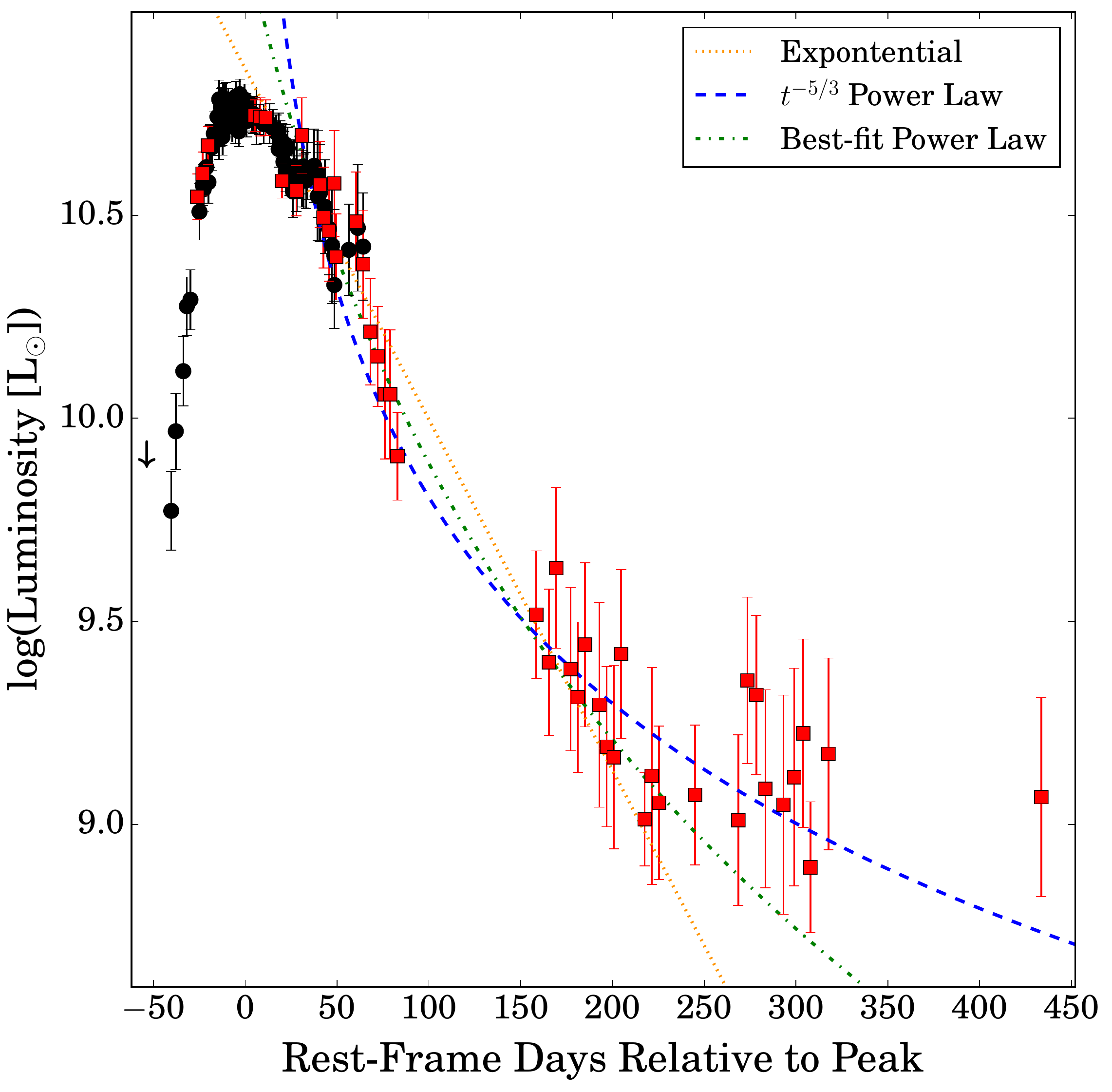}}}
\subfloat{{\includegraphics[width=0.48\textwidth]{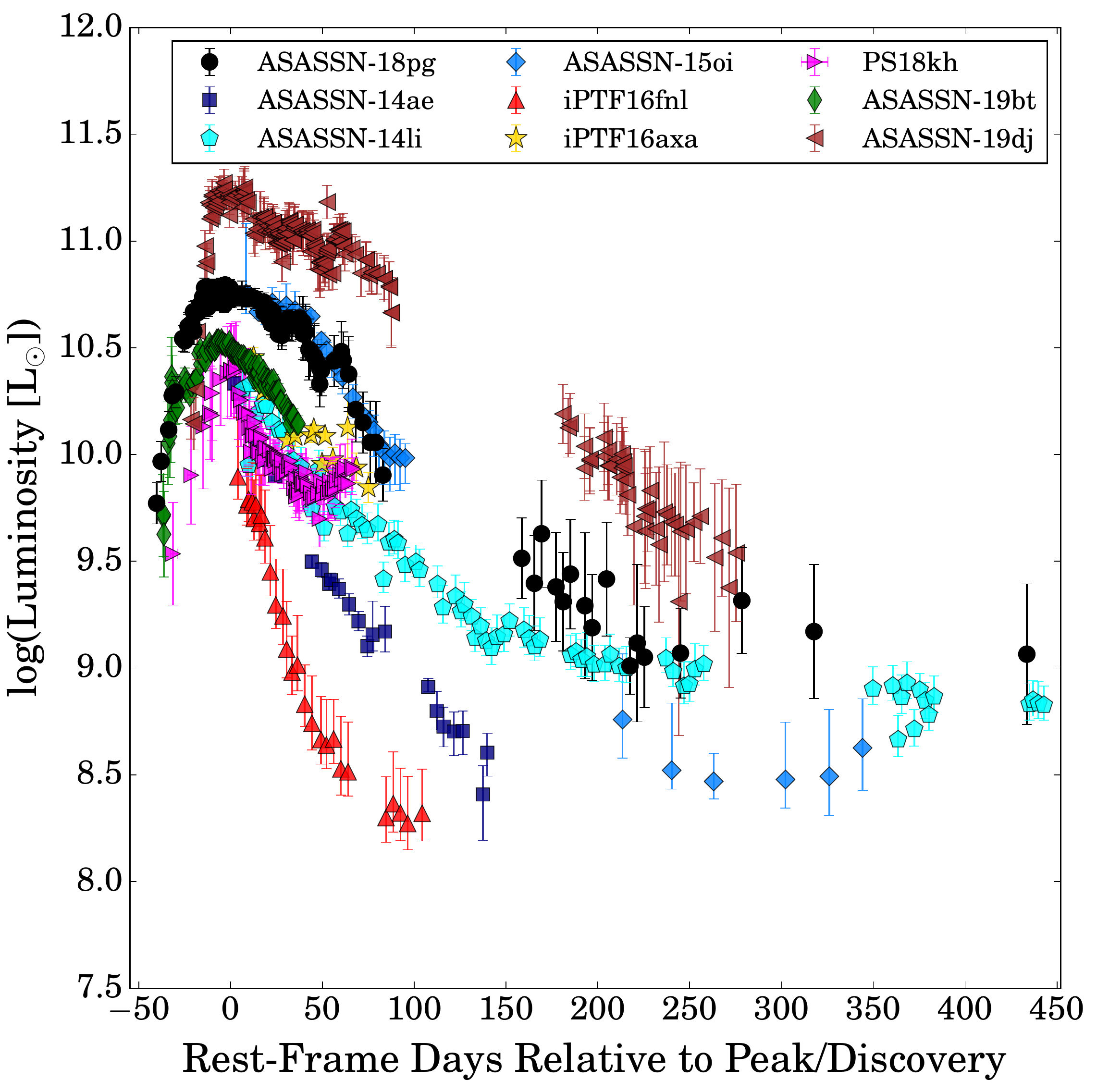}}}
\caption{\emph{Left Panel}: Evolution of the UV/optical luminosity of ASASSN-18pg from blackbody fits to the {\swift} SED (red squares) and $g$-band data that has been bolometrically corrected using the {\swift} fits (black circles). Downward arrows indicate upper limits. The yellow dotted, blue dashed, and green dash-dotted lines show exponential, $t^{-5/3}$ power-law, and $t^{-\alpha}$ power-law fits to the declining light curve, respectively. \emph{Right Panel}: Comparison of the luminosity evolution of ASASSN-18pg (black circles) to the luminosity evolutions of the TDEs ASASSN-14ae \citep[navy squares;][]{holoien14b}, ASASSN-14li \citep[cyan penatgons;][]{holoien16a}, ASASSN-15oi \citep[blue diamonds;][]{holoien16b}, iPTF16fnl \citep[red triangles;][]{brown18}, iPTF16axa \citep[gold stars;][]{hung17}, PS18kh \citep[magenta right-facing triangles;][]{holoien18a}, ASASSN-19bt \citep[green diamonds;][]{holoien19c}, and ASASSN-19dj (brown left-facing triangles; Hinkle et al., in prep.). Time is given in rest-frame days relative to peak for TDEs which have observations spanning the peak of the light curve (ASASSN-18pg, ASASSN-19bt, ASASSN-19dj, PS18kh, and iPTF16fnl) and in rest-frame days relative to discovery for TDEs which do not (ASASSN-14ae, ASASSN-14li, ASASSN-15oi, and iPTF16axa).}
\label{fig:lum_evol}
\end{minipage}
\end{figure*}

To get a better picture of the overall evolution of the bolometric luminosity, and to leverage the high-cadence light curves from ASAS-SN, Swope, and the Las Cumbres Observatory 1-m telescopes, we calculated bolometric corrections to the $g$-band light curve by linearly interpolating between the previous and next $g$-band measurements bracketing each \swift{} observation. We then used these bolometric corrections to estimate the bolometric luminosity of ASASSN-18pg from the full $g$-band light curve by linearly interpolating the bolometric corrections calculated for each \swift{} epoch to each epoch of $g$-band data. We used the bolometric correction from the first epoch of \swift{} SED fits to correct data taken prior to the first \swift{} observation. The full bolometric luminosity evolution calculated from the SED fits and the bolometrically corrected $g$ data is shown in Figure~\ref{fig:lum_evol}.

We fit the declining bolometric light curve ($t>30$~days after peak) with several profiles that have been used to fit declining TDE light curves in the past \citep[e.g.,][]{holoien19b}, including an exponential profile $L= L_0e^{-(t-t_0)/\tau}$, a $L=L_0 (t-t_0)^{-5/3}$ power-law profile, and a power law where the power-law index is allowed to vary, $L\propto (t-t_0)^{-\alpha}$. For the exponential profile we obtain best-fit parameters of $L_0=10^{44.6}$~erg~s$^{-1}$, $t_0=58323.3$, and $\tau=51.2$ days; for the $t^{-5/3}$ power law we obtain $L_0=10^{46.7}$~erg~s$^{-1}$ and $t_0=58345.2$; and for the free power law we obtain $L_0=10^{51.2}$~erg~s$^{-1}$, $t_0=58268.5$, and $\alpha=3.5$. All three fits are shown in Figure~\ref{fig:lum_evol}. 

The free power law provides the best fit, with $\chi^2_\nu=0.37$, and the exponential profile fits the data marginally better than the $t^{-5/3}$ power law, with $\chi^2_\nu=0.60$ compared to $\chi^2_nu=0.63$. The parameters of the exponential and $t^{-5/3}$ profiles are similar to those of other TDEs fit with the same procedure \citep[e.g.,][]{holoien19b}, but the parameters of the free power law are quite different, with the power law being significantly steeper. Despite having better $\chi^2$ than the $t^{-5/3}$ profile, however, neither the exponential nor the free power-law profiles fit the late-time data well, and even the $t^{-5/3}$ power law underestimates the luminosity in the latest epochs. Recent theoretical work predicts that there might be a transition in the dominant emission mechanism during TDE flares, with early, fallback-dominated emission following a steeper decline and later disk-dominated emission following a shallower power-law decline \citep[e.g.,][]{lodato11,auchettl17}. It is clear that none of the single models shown in Figure~\ref{fig:lum_evol} can fit the entire declining period perfectly, implying multiple physical processes are likely contributing to the observed emission. However, the $t^{-5/3}$ profile does fairly well and the best-fit $t_0$ is very close to our estimated peak date, which suggests that the emission from ASASSN-18pg may be largely fallback-dominated during the duration of our observations.

The right panel of Figure~\ref{fig:lum_evol} shows the luminosity evolution of ASASSN-18pg compared to several other TDEs from literature: ASASSN-14ae \citep{holoien14b}, ASASSN-14li \citep{holoien16a}, ASASSN-15oi \citep{holoien16b,holoien18a}, iPTF16fnl \citep{brown18}, iPTF16axa \citep{hung17}, PS18kh \citep{holoien19b}, ASASSN-19bt \citep{holoien19c}, and ASASSN-19dj (Hinkle et al., in prep.). The rise of ASASSN-18pg looks generally similar to those of ASASSN-19bt and PS18kh, though it lacks the early luminosity spike seen before peak in ASASSN-19bt \citep{holoien19c}. With a peak luminosity of $L_{\textrm{peak}}\simeq2.2\times10^{44}$~erg~s$^{-1}$, ASASSN-18pg is one of the most luminous TDEs in the sample, and it exhibits a period of slower decline following peak that looks very similar to those of ASASSN-15oi and ASASSN-19dj, both of which are also quite luminous. This is consistent with the general finding by \citet{hinkle20a} that more luminous TDEs decline more slowly after peak.


\begin{figure}
\centering
\includegraphics[width=0.425\textwidth]{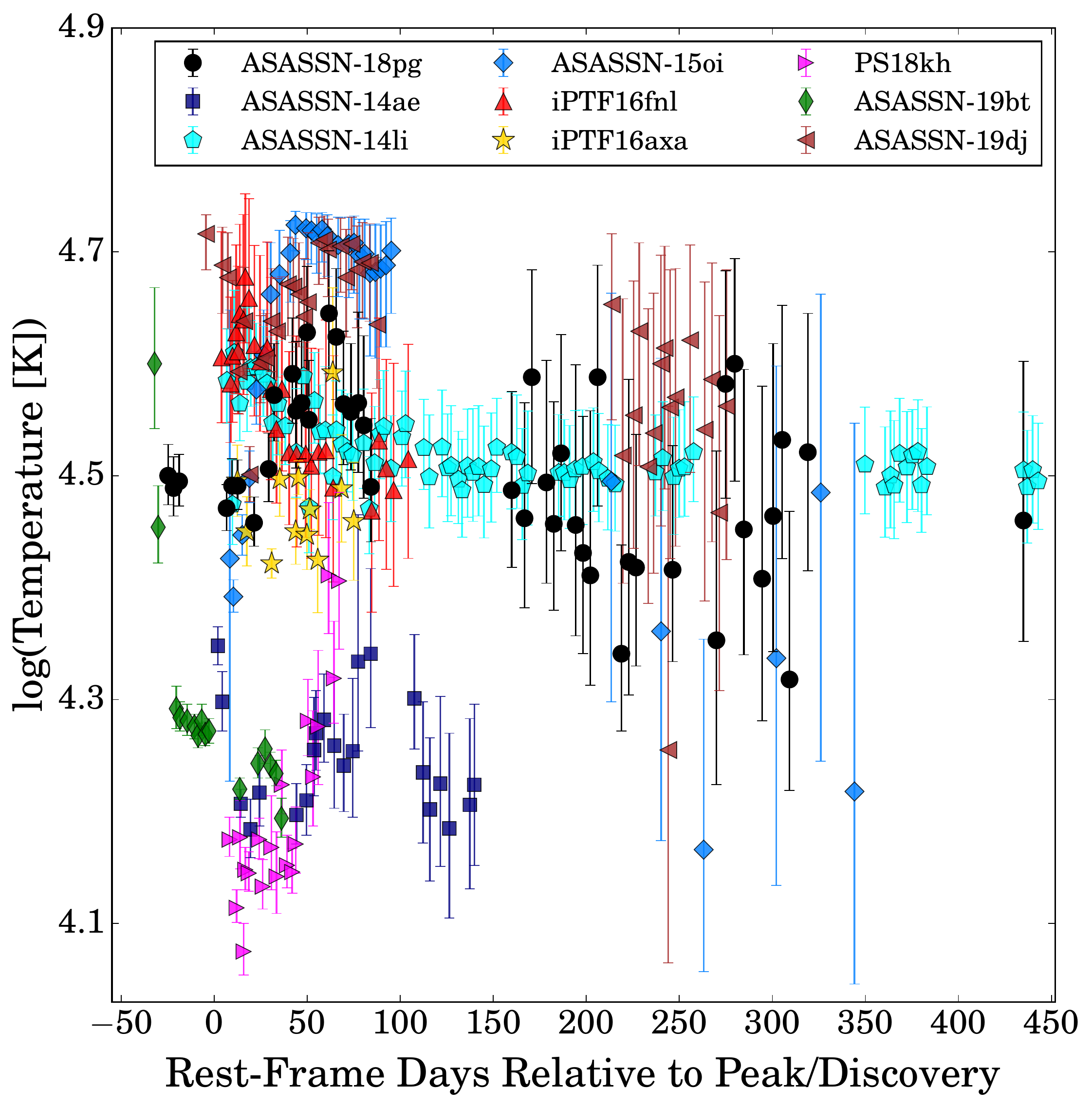}
\caption{Temperature evolution of ASASSN-18pg from our {\swift} blackbody fits (black circles) compared to those of the TDE comparison sample. As described in the caption of Figure~\ref{fig:lum_evol}, time is shown in rest-frame days relative to peak or discovery, and symbols and colors match those of Figure~\ref{fig:lum_evol}.}
\label{fig:temp_evol}
\end{figure}

Integrating over the rest-frame bolometric light curve, ASASSN-18pg radiated a total of $E=(1.78\pm0.05)\times10^{51}$~ergs, with roughly 30\% ($(5.38\pm0.09)\times10^{50}$~ergs) radiated during the rise to peak. This is significantly more energy output than other recent TDEs \citep[e.g.,][]{holoien19b,holoien19c}, which is not a surprise given the relatively high luminosity and slow post-peak decline exhibited by ASASSN-18pg. An accreted mass of $M_{Acc}\simeq0.009\eta_{0.1}^{-1}$~\msun, where the accretion efficiency is $\eta=0.1\eta_{0.1}$, is required to generate the emitted energy. This is very low compared to the mass estimate of the star, as has been seen in other TDEs. ASASSN-18pg thus once again indicates that it is likely only a small fraction of the stellar material actually accretes onto the SMBH during a TDE, or that the radiative efficiency is quite low \citep[e.g.,][]{holoien14b,holoien16a,holoien18a}.

The blackbody temperature evolution of ASASSN-18pg from the \swift{} fits is shown in Figure~\ref{fig:temp_evol} along with the evolution for the same comparison sample shown in Figure~~\ref{fig:lum_evol}. ASASSN-18pg shows very little temperature evolution throughout the duration of the flare, remaining roughly constant around $T\sim30,000$~K until late times. There is some evidence of a temperature increase/spike around 60 rest-frame days after peak, possibly reaching as high as $T\sim45,000$~K, but the uncertainties are large enough that this spike may not be quite so dramatic. It is clear that ASASSN-18pg does not exhibit any of the more dramatic changes seen in some of the other TDEs, such as the early temperature drop of ASASSN-19bt \citep{holoien19c}, the early rises of ASASSN-15oi and PS18kh \citep{holoien18a,holoien19b}, or the late-time drop of ASASSN-15oi \citep{holoien18a}. Our \swift{} observations of ASASSN-18pg cover a long enough time baseline to make comparison at both very early and late times possible, which has not been the case with any other TDE in the sample. The lack of an early drop in the temperature as seen in ASASSN-19bt is of note, as ASASSN-19bt was the first TDE with UV data to fit the blackbody temperature at such early times, and it is unclear how common such an early temperature decline is.


\begin{figure}
\centering
\includegraphics[width=0.425\textwidth]{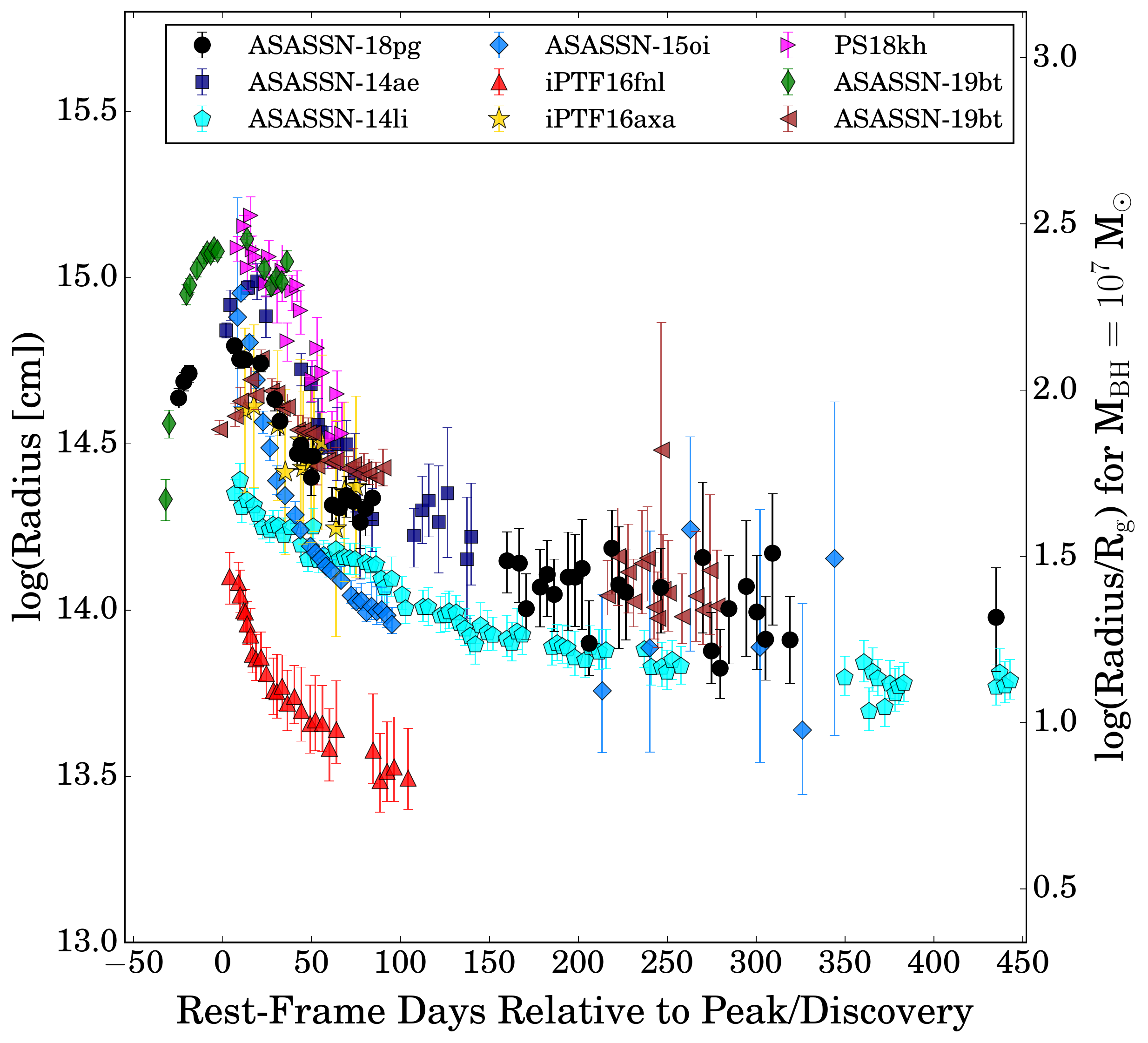}
\caption{Evolution of the blackbody radius of ASASSN-18pg from the {\swift} fits compared to those of the TDEs in the comparison sample. As described in the caption of Figure~\ref{fig:lum_evol}, time is shown in rest-frame days relative to peak or discovery, and symbols and colors match those of Figure~\ref{fig:lum_evol}. The scale on the left shows the radius in units of cm and the scale on the right shows the same scale in units of the gravitational radius for a $10^7$~\msun~black hole.}
\label{fig:rad_evol}
\end{figure}


\begin{figure*}
\begin{minipage}{\textwidth}
\centering
\subfloat{{\includegraphics[width=0.32\textwidth]{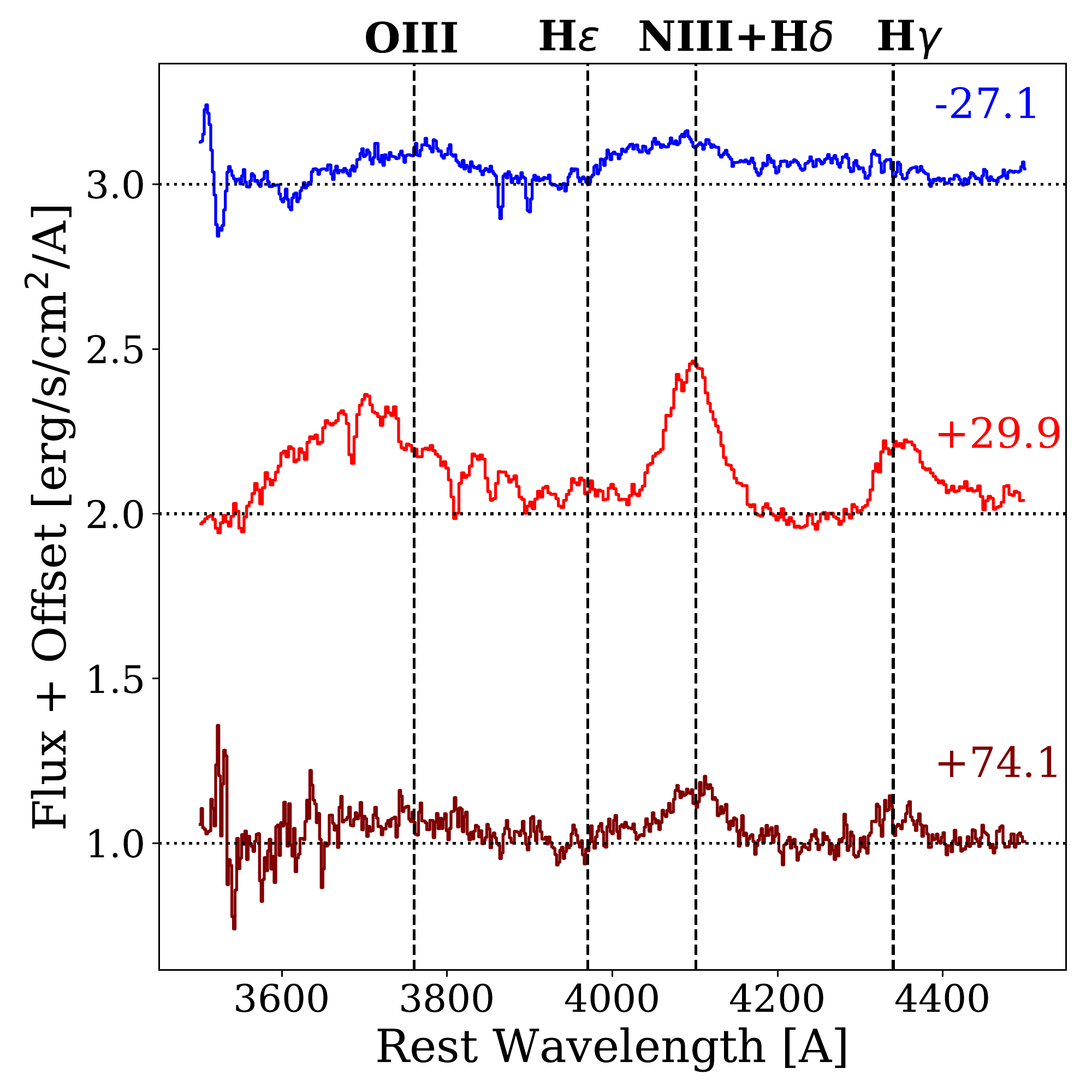}}}
\subfloat{{\includegraphics[width=0.32\textwidth]{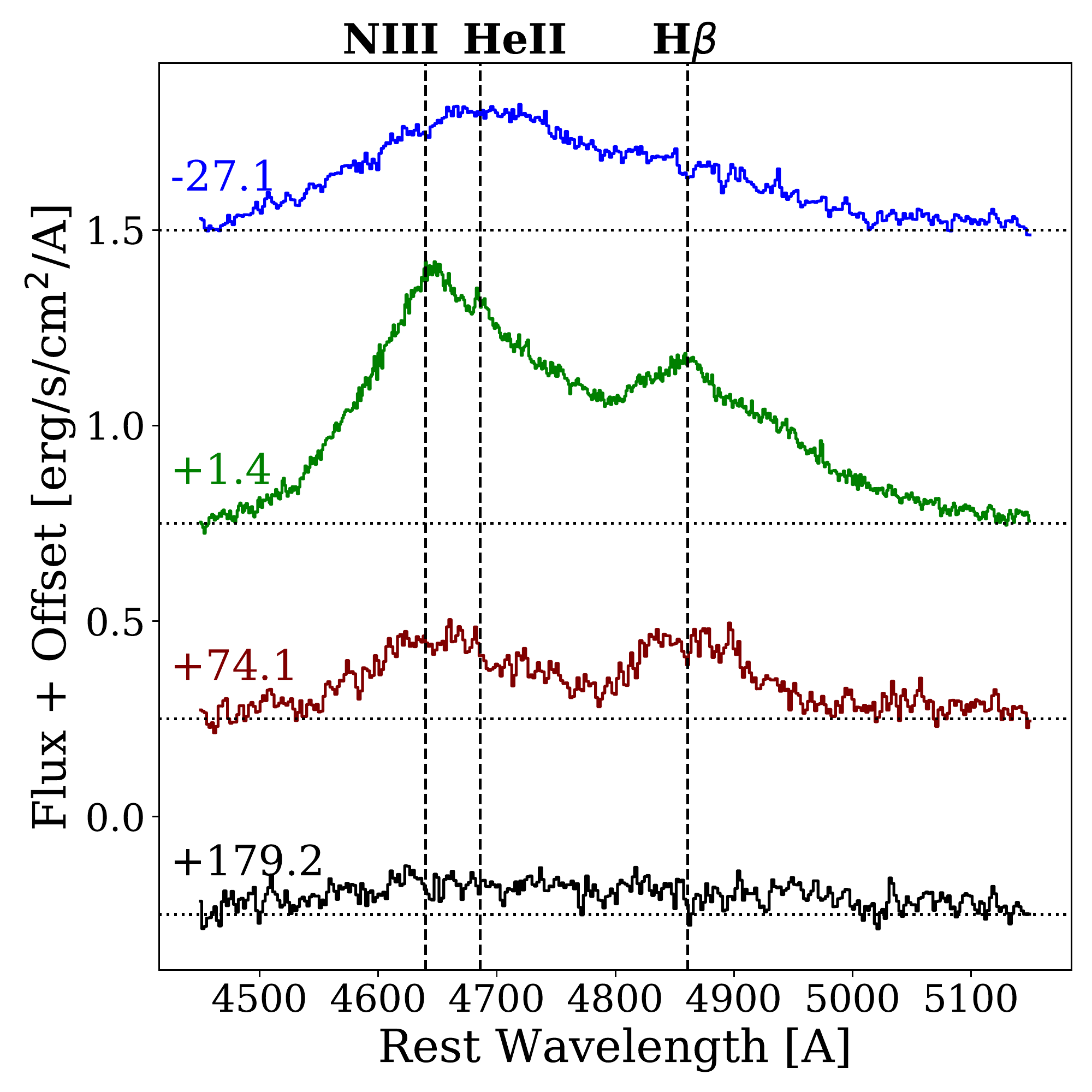}}}
\subfloat{{\includegraphics[width=0.32\textwidth]{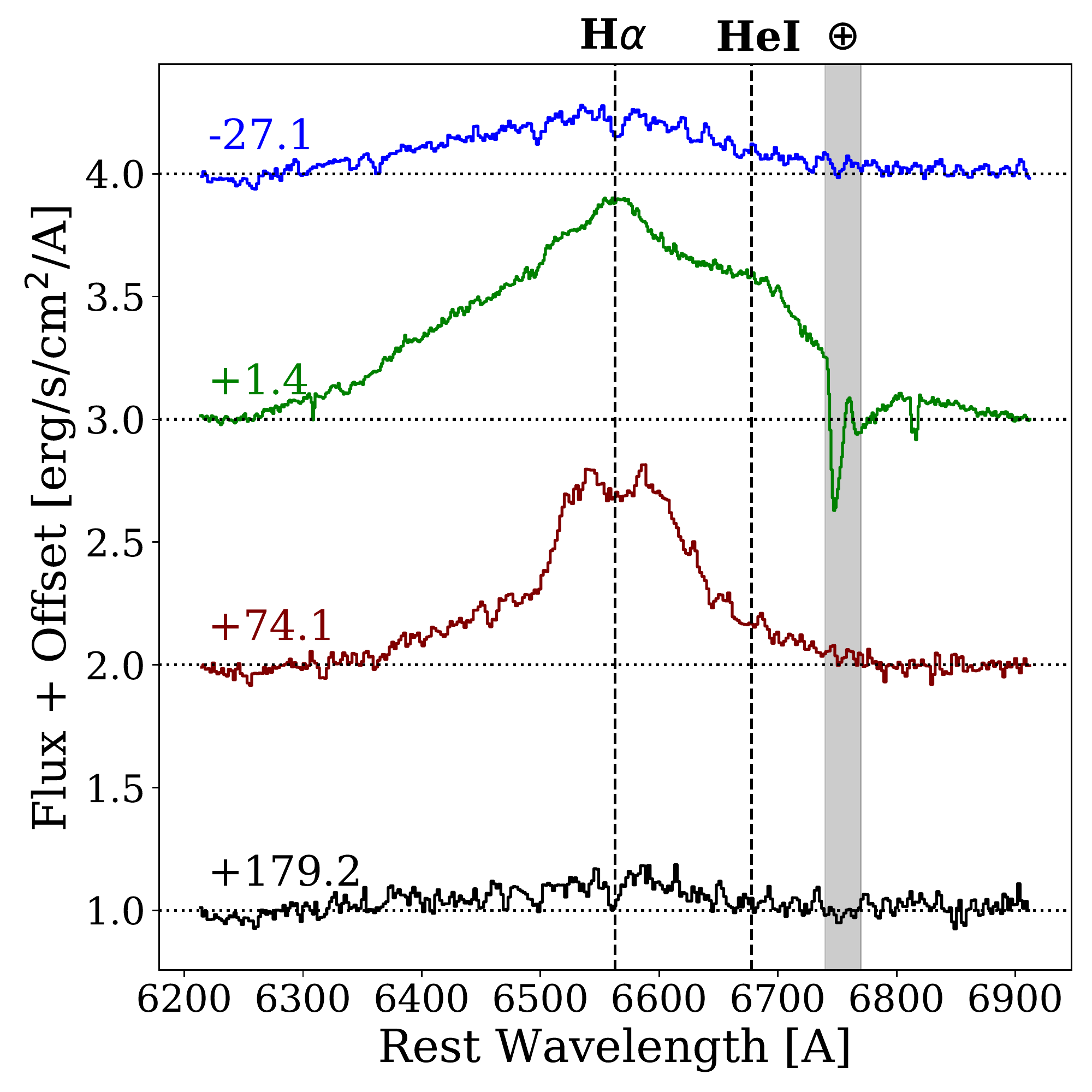}}}
\caption{Evolution of the spectroscopic emission features centered on the \ion{O}{3} 3760\AA{} triplet and 4100\AA{} line complex (left panel), \hbeta{} (center panel), and \halpha{} (right panel). Spectra were chosen to show the lines roughly one month prior to peak, near peak, roughly two months after peak shortly before it becomes Sun-constrained, and roughly six months after peak, with the epoch in rest-frame days relative to peak shown next to each spectrum. The spectra shown in the left panel differ from those of the other two panels in some cases, as only some of our follow-up spectra cover this wavelength range. Prominent lines are indicated with dashed lines and a linear continuum has been subtracted from each spectrum.}
\label{fig:line_evo}
\end{minipage}
\end{figure*}

Finally, in Figure~\ref{fig:rad_evol} we show the evolution of the blackbody radii of ASASSN-18pg and the comparison TDEs. Similar to ASASSN-19bt, the radius increases prior to peak light in ASASSN-18pg, though the rise is considerably slower than that of ASASSN-19bt. Following peak, the radius declines fairly rapidly before leveling off at later times, and appears to be very similar in size and evolution to those of ASASSN-19dj and ASASSN-14ae in particular. In general, the radius evolution of TDEs appears to be much more homogeneous than the temperature or luminosity evolutions, with the majority of the objects in our sample exhibiting similar sizes and evolutions. In general it seems that TDEs with hotter temperatures have smaller peak radii, but there does not appear to be much difference in the rate of change of the radius with different temperatures. However, few of the TDEs in this sample have both early- and late-time data, making it difficult to draw conclusions about potential trends, particularly past $\sim100$ days post-peak.

\section{Spectroscopic Analysis}
\label{sec:spec_anal}

\subsection{Emission Line Analysis}
\label{sec:spec_line_anal}

As noted by \citet{leloudas19}, ASASSN-18pg is a member of a new class of TDEs that exhibit several emission lines resulting from the Bowen flourescence mechanism in addition to the broad hydrogen and helium lines common to TDEs. Several other TDEs with similar features have now been identified \citep{blagorodnova19,velzen20}. Many of these lines are clearly detected in several of our spectra, and we fit these features in order to measure the evolution of the lines. 


\begin{figure*}
\begin{minipage}{\textwidth}
\centering
\subfloat{{\includegraphics[width=0.32\textwidth]{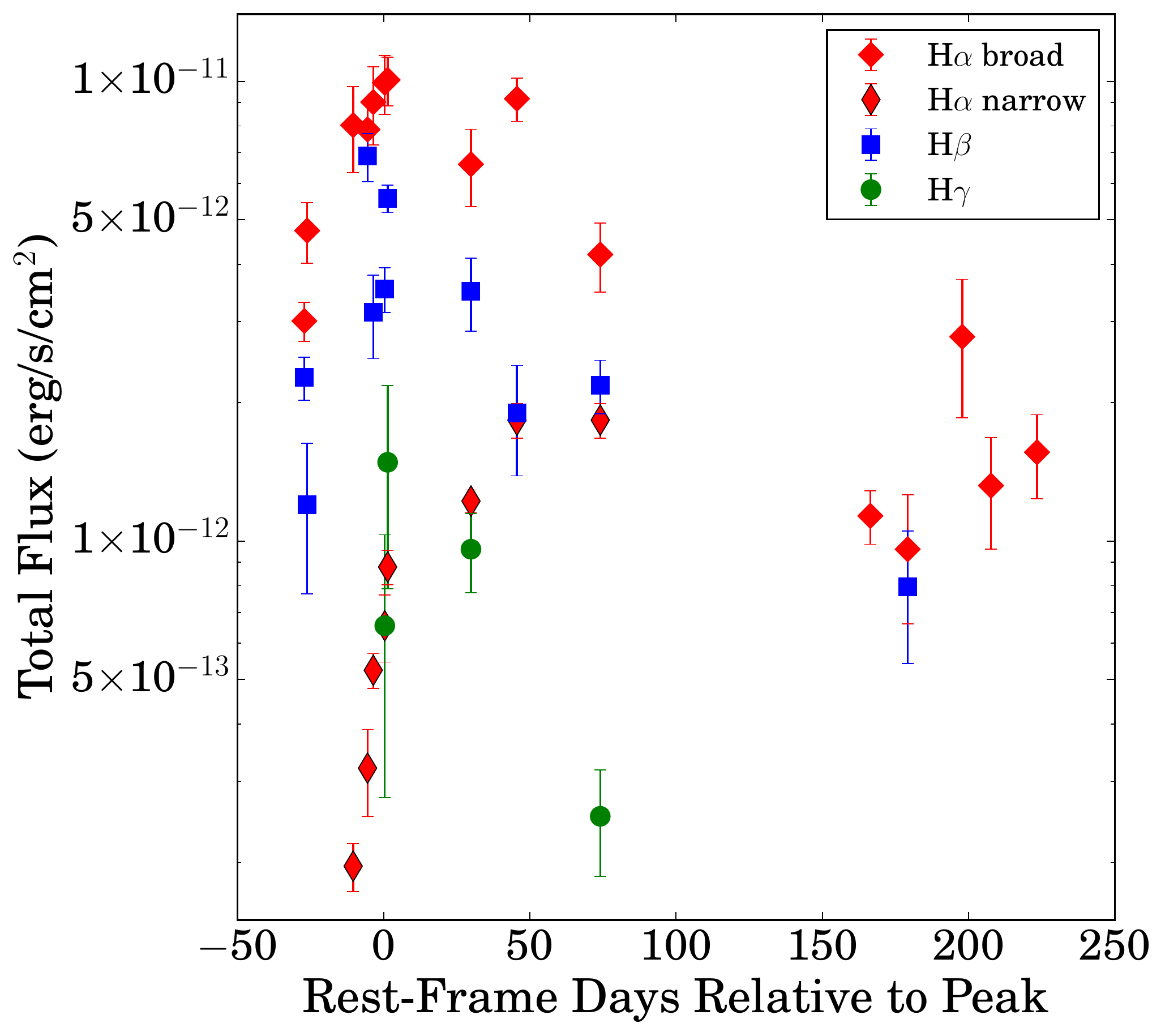}}}
\subfloat{{\includegraphics[width=0.32\textwidth]{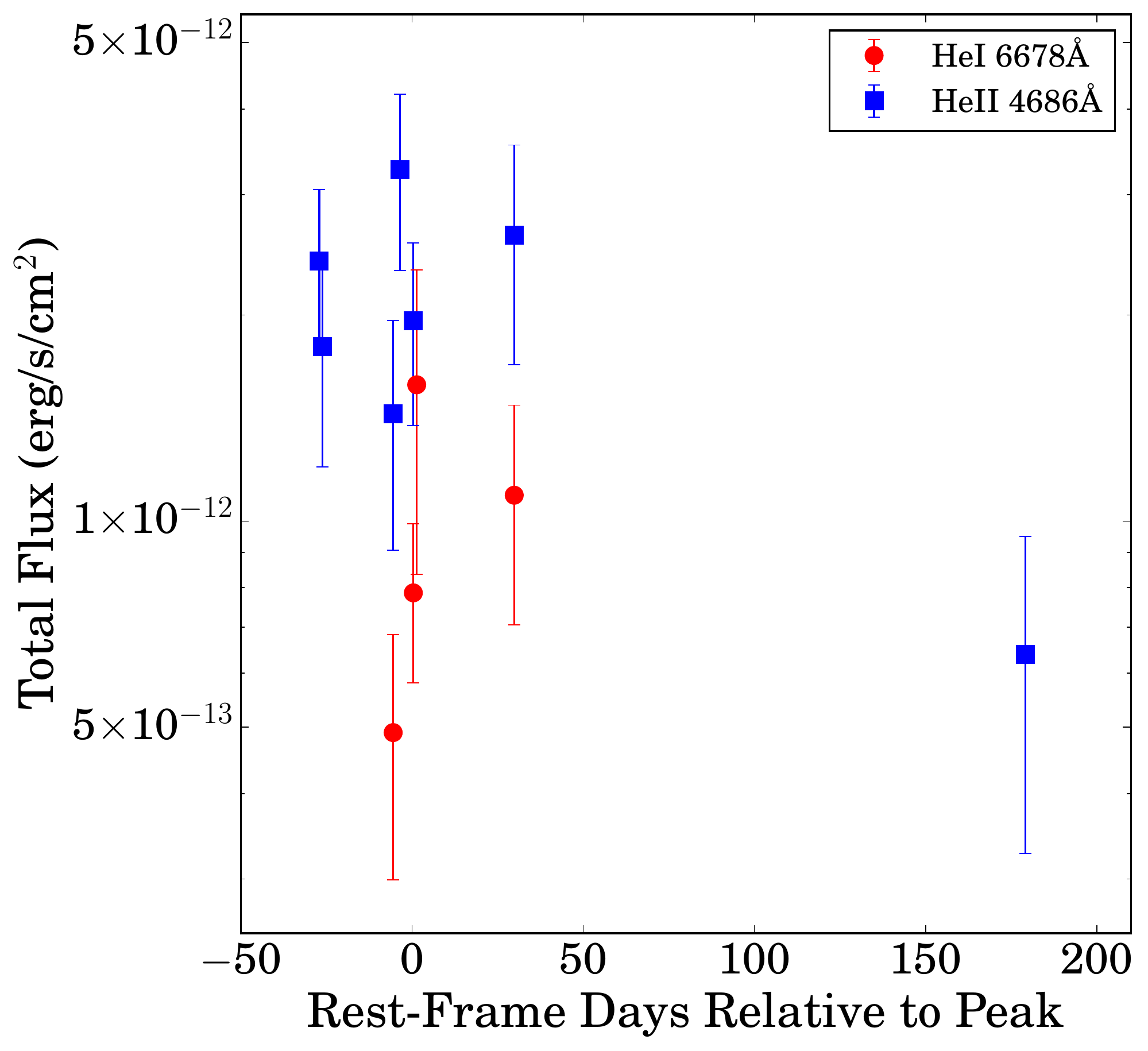}}}
\subfloat{{\includegraphics[width=0.32\textwidth]{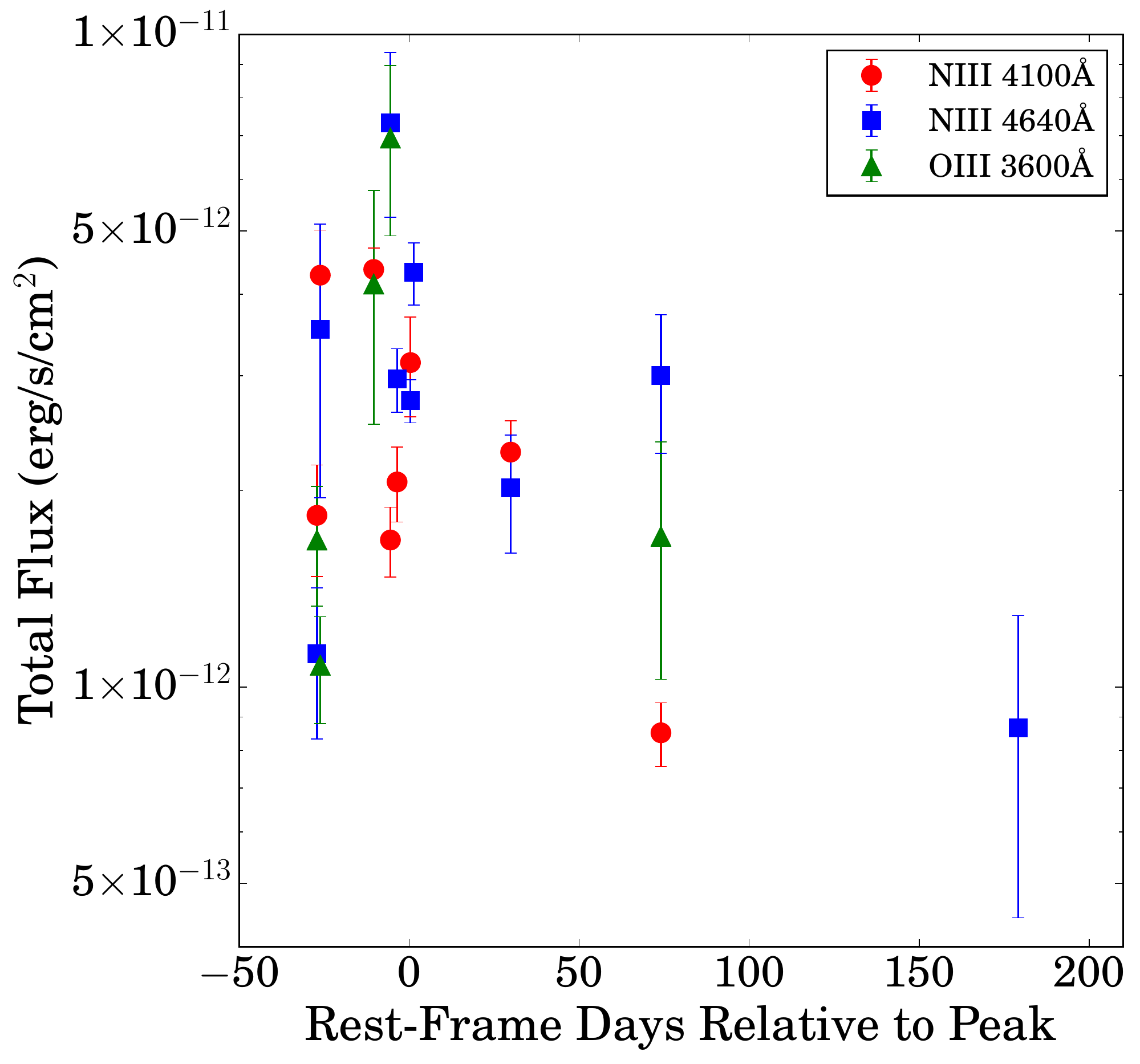}}}
\caption{\emph{Left Panel}: Evolution of the fluxes of the \halpha{} broad and narrow components (wide and thin red diamonds, respectively), the \hbeta{} line (blue squares), and the \hgamma{} line (green circles). \emph{Center Panel}: Evolution of the fluxes of the \ion{He}{2} 4686\AA{} (red circles) and \ion{He}{1} 6678\AA{} (blue squares) lines. \emph{Right Panel}: Evolution of the fluxes of the \ion{N}{3} 4100\AA{} (red circles), \ion{N}{3} 4640\AA{}(blue squares), and \ion{O}{3} 3600\AA{} (green triangles) lines.}
\label{fig:line_fluxes}
\end{minipage}
\end{figure*}

We fit the lines as Gaussian profiles atop a linear continuum. Due to the broadness of the emission profiles and the low S/N of most of our spectra, we manually selected regions of the spectrum near each line for continuum estimation. The continuum was removed and the remaining emission profiles were fit with three free parameters: velocity width, velocity shift from the rest wavelength, and amplitude of the emission profile. The parameters were initially estimated by manually adjusting the values until a reasonable fit was achieved, followed by using a least-squares minimizer to fine-tune the results. Due to the number of broad emission lines, identifying and removing the proper continuum level is non-trivial and likely contributes $\gtrsim25$\% of our overall error budget. For this reason, we focus more on relative changes as ASASSN-18pg evolves, rather than absolute measurements. We only attempt to fit the lines in our spectra taken through the end of 2019 March, as no transient emission line features are detected in our later observations.

Due to the heavy telluric contamination between 6830\AA{} and 6980\AA{} (6710\AA{}$-$6860\AA{} rest-frame), decomposing the individual contributions from \halpha{} and \ion{He}{1}~6678\AA{} is non-trivial. When ASASSN-18pg is near peak light, the \halpha{} and \ion{He}{1} emission lines are strong enough to provide a general estimate of the properties for both lines (e.g, Figure~\ref{fig:line_evo}, green spectrum), albeit with large uncertainties. We provide the flux estimates for the \ion{He}{1}~6678\AA{} line in the middle panel of Figure~\ref{fig:line_fluxes}, but caution that the uncertain continuum level is a large source of systematic uncertainty. For all other epochs we only fit the peak and left-wing of the \halpha{} profile to estimate emission-line properties.

The \halpha{} feature starts out as a single, broad Gaussian with width of $\sim15,000$~km~s$^{-1}$ in our early spectra. Over time, a narrow peak (width of $\sim6,000-7,000$~km~s$^{-1}$) develops atop the broad \halpha{}. The narrow feature becomes dominant as time progresses, but has disappeared after ASASSN-18pg reappeared from being sun-constrained. After the narrow feature has appeared, we fit the narrow and broad components simultaneously with Gaussian profiles, as the narrow emission component is narrow enough where the telluric absorption is not an issue. 

Overall, our results are consistent with those found by \citet{leloudas19}, with the exception of the two-component \halpha{} line. No emission lines show any significant deviation from the rest wavelength. The evolution of several continuum-subtracted emission features are shown in Figure~\ref{fig:line_evo} and described below, and we show the evolution of the fluxes of the various lines in Figure~\ref{fig:line_fluxes}.

The broad \halpha{} component grows broader from 2018 July until 2018 August, with the FWHM increasing from $\sim12,000$~km~s$^{-1}$ to $\sim15,000$~km~s$^{-1}$. The narrow \halpha{} component becomes visible in 2018 August, growing stronger over time and becoming roughly equivalent in strength to the broad component shortly before ASASSN-18pg becomes Sun-constrained in 2018 November. After the TDE has re-emerged from behind the Sun in 2019 January, the broad component is still detected with $\textrm{FWHM}\sim6,000$~km s$^{-1}$, but the narrow component is no longer detected. The broad \halpha{} component becomes fainter over time with similar FWHM, and is no longer detected after 2019 March.

Broad \hbeta{} emission is seen throughout the evolution of the TDE, with the line narrowing from $\textrm{FWHM}\sim12,000-15,000$~km~s$^{-1}$ in 2018 July and August to $\textrm{FWHM}\sim6,000-7,000$~km~s$^{-1}$ in 2018 September and October. After re-emerging from being Sun constrained, the \hbeta{} region is dominated by a complex of lines also including \ion{He}{2}~4686\AA{} and \ion{N}{3}~4640\AA{} lines, and it is difficult to extract each line's individual contributions. \hbeta{} likely continues to be detected until late 2019 March. We do not detect a narrow \hbeta{} line similar to the narrow \halpha{} component in any epoch.

The \hgamma{} line is undetected until the spectrum obtained on 2018 August 14, where we tentatively detect weak \hgamma{} emission with $\textrm{FWHM}\sim8,000$~km~s$^{-1}$. It remains detected until the TDE became Sun-constrained, with the emission peaking in strength on 2018 September 13, and is not detected after. The only plausible detection of \hdelta{} occurs on 2018 September 13, when the \hgamma{} emission is strongest, corresponding to a shoulder on the red wing of the \ion{N}{3}~4100\AA{} emission profile (see below).

Weak \ion{He}{1}~6678\AA{} and \ion{He}{2}~4686\AA{} lines become visible in 2018 August and are blended with the \halpha{} and \hbeta{} lines, respectively. The \ion{He}{2}~4686\AA{} line is particularly weak compared to the \hbeta{} and \ion{N}{3}~4640\AA{} lines in the same region of the spectra. \ion{He}{1}~6678\AA{} is not detected after the TDE becomes visible again in 2019, and the \ion{He}{2}~4686\AA{} line is only tentatively detected once at later times.

Similar to \citet{leloudas19}, we detect a feature near 5800~\AA, which was speculated to be a blend of \ion{He}{1}~5876\AA{} and [\ion{N}{2}]~5754\AA{} lines. This region cannot be fit by any reasonable combination of these two line profiles for any of our spectra unless these lines have significant shifts from their rest wavelengths not evident in any of the other emission lines. Significant host galaxy and/or Milky Way Na ID absorption complicates the fitting process. Thus, we conclude that the origin of this emission feature remains ambiguous.

Finally, \citet{leloudas19} identified several emission lines seen from ASASSN-18pg as the result of Bowen flourescence, and \citet{velzen20} have since discovered several other TDEs with similar emission features. We also detect many of these lines in our spectra at various times. The \ion{N}{3}~4640\AA{} is of similar width and flux to the \hbeta{} line in most epochs, evolving similarly to become stronger and broader for roughly the first month after discovery and remaining detected at late times. 

In agreement with \citet{leloudas19}, we clearly detect the \ion{O}{3}~3760\AA{} triplet and an emission complex near 4100\AA{} that we consider likely to be \ion{N}{3}~4100\AA{} emission, rather than \hdelta. The evolution of the \ion{O}{3}~3760\AA{} line roughly tracks that of \ion{N}{3}~4640\AA, while the \ion{N}{3} blend begins broad and gradually decreases in aplitude and FWHM over time.

We observe an apparent delay between the times when the broad and narrow \halpha{} components peak. Motivated by this, we examined whether these components could illuminate the geometry of the gas responsible for the spectroscopic features. To do this, we first took the bolometric luminosity curve and extrapolated to times earlier than our first detection as a power law. We then convolved this bolometric curve with the transfer functions of spherical shells of several radii, treating the delay distribution of each shell as a top hat running from a delay of zero to $2R_{shell}/c$. This produced light curves one might expect for the simplistic case of a spherical shell of gas reprocessing some of the UV emission from the TDE into recombination line emission.

\begin{figure}
\centering
\includegraphics[width=0.48\textwidth]{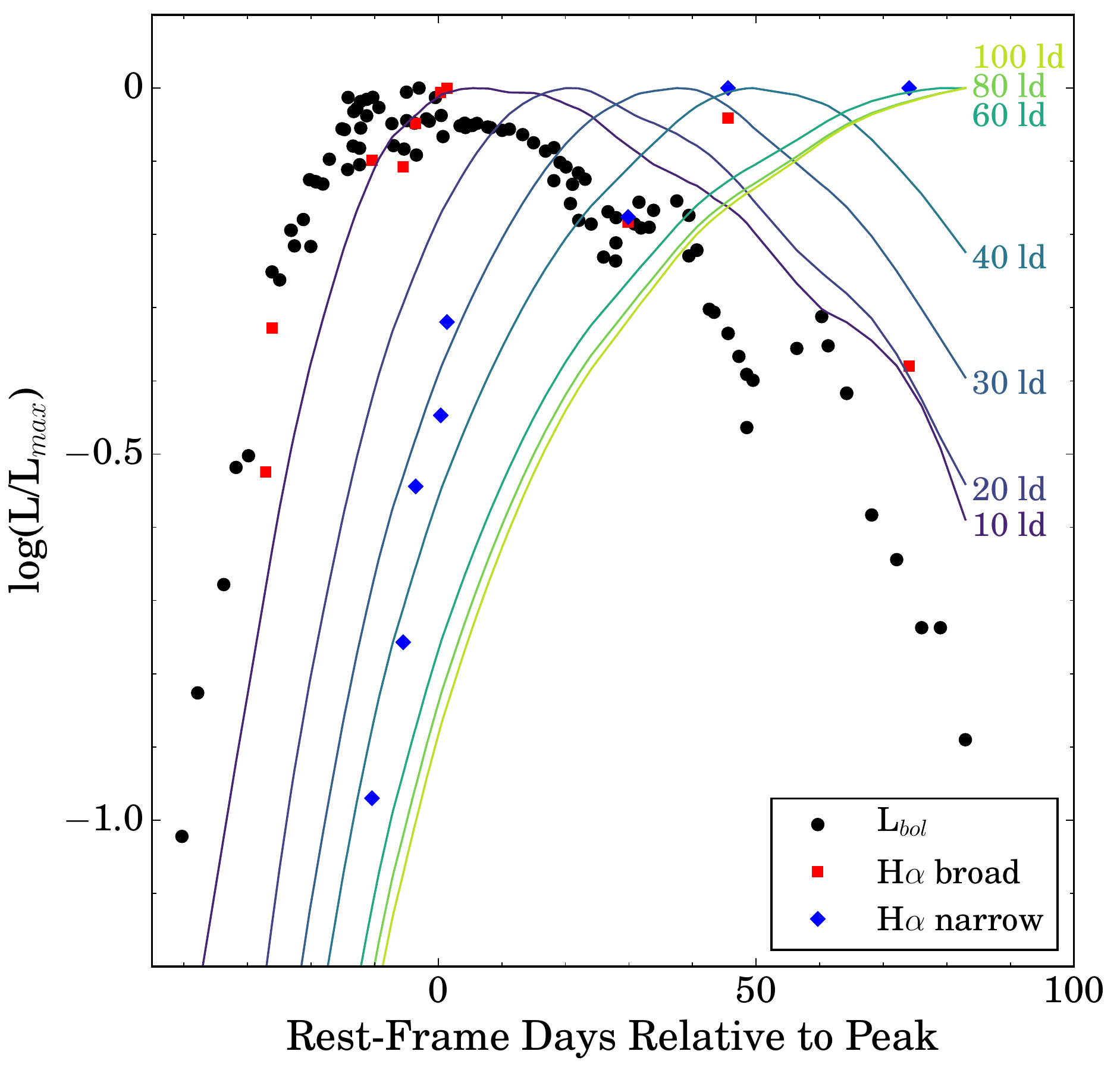}
\caption{Light curves of the bolometric luminosity (black circles), broad \halpha{} component (red squares), and narrow \halpha{} component (blue diamonds), normalized to the peak luminosity for each. The colored lines show the expected luminosity curves for spherical shells of different radii reprocessing the bolometric emission, with the radius in light days given next to each. The bolometric light curve prior to the first detection was extrapolated as a power law to earlier times, and the delay distribution of each shell is a top hat running from a delay of zero to $2R_{shell}/c$.}
\label{fig:line_lag}
\end{figure}

In Figure~\ref{fig:line_lag} we show the luminosities of the bolometric, broad \halpha, and narrow \halpha{} components, each normalized to their maximum values. Overplotted on the data are the spherical shell approximations, running from 10 light days to 100 light days. Though none of the extrapolated curves fit the spectroscopic components exactly, the broad \halpha{} luminosity is roughly consistent with a shell of $10-20$ light days, while the narrow component is roughly consistent with a shell of $40-60$ light days. The geometry of reprocessing gas is likely much more complicated than a simple spherical shell, and is also likely evolving an fairly rapid timescales, but the fact that these simple approximations can reasonably fit the data implies that these spectroscopic features are likely coming from different regions around the black hole, with the narrow component likely being significantly more distant and slower moving than the broad component. 

We note that we detect broad emission features in all spectra obtained prior to peak, including our first spectrum obtained approximately $27$ rest-frame days before peak. This is in contrast to several other recent TDEs with spectroscopic observations at similar times, such as PS18kh, ASASSN-19bt, and ASASSN-19dj, which exhibited a strong blue continuum but no emission features until closer to peak light \citep[][Hinkle et al., in prep.]{holoien19b,holoien19c}. There are roughly 16 rest-frame days between our first detection of the TDE in ASAS-SN data and our first spectrum, so we cannot rule out the possibility that the emission features would not be detected if ASASSN-18pg had been observed earlier. However, if the lines are present in all epochs, this perhaps suggests a different physical origin for these features in ASASSN-18pg than in other TDEs with early observations. In particular, ASASSN-18pg is a Bowen TDE, while PS18kh, ASASSN-19bt, and ASASSN-19dj are all H-rich TDEs. As Bowen flourescence is a process that requires reprocessing of higher energy emission, this suggests that the emission features in ASASSN-18pg are driven by reprocessing of emission from the accretion disk, and that the lines are present in all epochs because the UV/optical emission is not detected until the disk emission has been reprocessed \citep[e.g.,][]{roth16,roth18}. If the UV/optical emission in the other TDEs is driven by shocks in the tidal debris stream as it collides with itself \citep[e.g.,][]{piran15,krolik16}, it is possible we may not observe lines until later times, or see more rapid variation, as the material responsible for the emission is evolving on short timescales. Viewing angle may also play an important role in the observed difference between ASASSN-18pg and these other objects \citep[e.g.,][]{dai18}. More TDEs with very early-time spectroscopic observations such as these are needed to determine if there truly is a subset of TDEs that exhibit lines in all epochs, and to determine the origin of the different timescales we observe in the emergence of the emission features in TDEs.

\subsection{Spectropolarimetry of ASASSN-18pg}
\label{sec:specpol}

\begin{figure}
\centering
\includegraphics[width=0.48\textwidth]{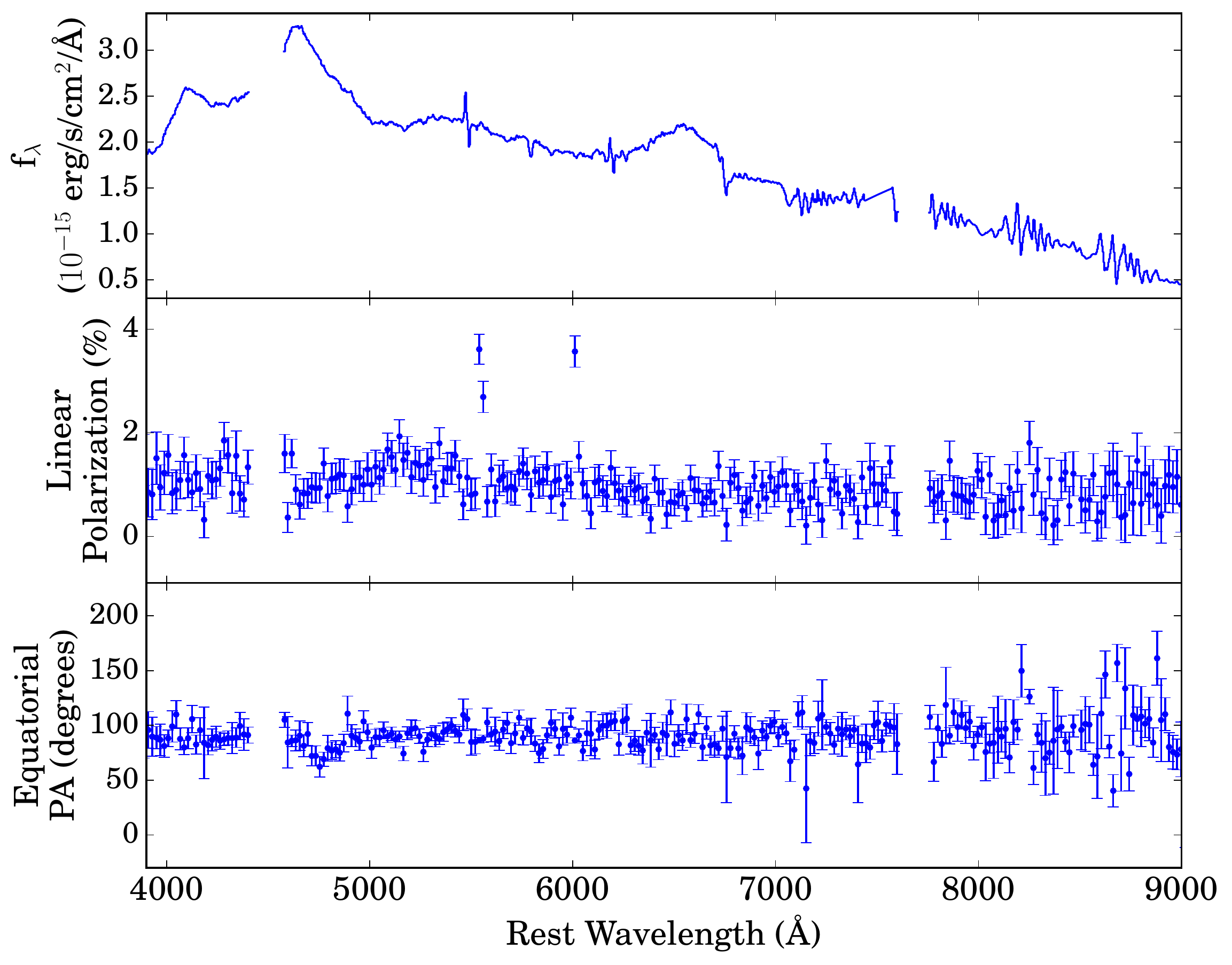}
\caption{Spectropolarimetric observation of ASASSN-18pg obtained on 2018 August 03 from SALT. \emph{Top Panel:} Photometry-calibrated spectrum, also shown in Figure~\ref{fig:spec_evol}. \emph{Middle panel:} Linear polarization. \emph{Bottom Panel:} Instrumental equatorial position angle. Both the linear polarization and the position angle have been binned in 20\AA{} bins to increase readability. We do not display observations at wavelengths shorter than 4000\AA{} or longer than 9000\AA, where uncertainties on the polarization and position angle are large.}
\label{fig:specpol}
\end{figure}

As mentioned in Section~\ref{sec:spec_obs}, the SALT spectrum obtained on 2018 August 03 was a low-resolution spectropolarimetric observation. Such observations can be useful for determining the geometry of the emission source and may be particularly illuminating for TDEs, which are expected to be quite aspherical and rapidly evolving \citep[e.g.,][]{guillochon13,guillochon15}, particularly at times shortly after disruption. These observations were obtained roughly 10 rest-frame days prior to peak, and represent the first published spectropolarimetric observations of a TDE. The spectrum, polarization, and instrumental position angle are shown in Figure~\ref{fig:specpol}.

We detect a nominal polarization of $\sim1.5$\% with uncertainties of $\sim0.5$\% that remains roughly constant with some slight variation from $4000-9000$\AA. We only examine this wavelength range, as uncertainties on the polarization and position angle balloon at shorter and longer wavelengths. We do not see any obvious features in the polarization associated with the emission lines as compared to the continuum, implying that the lines and continuum come from the same source with the same geometry.

In order to determine whether this polarization is intrinsic to the TDE, we first examine whether the polarization could be consistent with interstellar polarization (ISP), which is induced by dichroic absorption of the TDE light by interstellar dust grains aligned to the magnetic field of the interstellar medium (ISM) along the line of sight to the TDE. Due to the high Galactic extinction in the direction of the TDE ($E(B-V)\simeq0.2$~mag), there could be up to $P_{\textrm{ISP}}<9E(B-V)\simeq1.8$\% Galactic ISP \citep{serkowski75,bose18c}, which is consistent with the polarization we measure. Polarization measurements of three stars within 1 degree of ASASSN-18pg also support this, with the stars having polarization of $\sim1$\% at position angles of $\sim50$ degrees \citep{heiles00}, similar to what we measure in our observation of ASASSN-18pg. Thus, we conclude that the detected polarization is likely due to Galactic ISP.


However, if we assume the polarization is intrinsic to the TDE, a polarization of $2$\% corresponds to an axis ratio in the emission region of $\sim0.65$ and a polarization of $1$\% corresponds to an axis ratio of $\sim0.8$, assuming that the \citet{hoflich91} analysis for supernovae can be applied to TDEs. If the polarization is intrinsic to the TDE, these observations imply that the emission region cannot be highly aspherical. 

\section{Summary and Discussion}
\label{sec:disc}

With several hundred observations spanning from 54 days prior to peak light through 441 days after peak light, our data on ASASSN-18pg represent one of the most comprehensive early-through-late-time datasets available for a TDE. It includes X-ray, UV, optical, and radio observations as well as spectra spanning several hundred days of the TDE's evolution and the first published spectropolarimetric observations of a TDE.

Due to our early discovery and subsequent triggering of additional follow-up resources, our data include multiwavelength data prior to peak spanning from the X-ray through optical wavelengths as well as several spectra taken before and around peak light, allowing us to characterize the early blackbody evolution and spectroscopic evolution of this TDE. Our early photometry provide strong constraints on the rise time, in turn providing good constraints on the black hole mass, star mass, and viscous delays, parameters of the {\tt MOSFiT} model. The blackbody fits indicate that ASASSN-18pg peaked at a luminosity of $L_{\textrm{peak}}\simeq2.2\times10^{44}$~erg~s$^{-1}$, making it one of the more luminous UV/optical TDEs discovered to-date. It declines at a relatively slower rate than less luminous TDEs, and follows the peak luminosity-luminosity decline rate relation discovered by \citet{hinkle20a}. 

ASASSN-18pg is a member of the recently identified class of TDEs that exhibit emission lines attributed to Bowen flourescence \citep{leloudas19}. Unlike other TDEs which developed emission lines in their spectra at or shortly before peak light \citep[e.g.,][]{holoien19b,holoien19c}, ASASSN-18pg exhibits emission lines in all epochs, including our earliest spectrum obtained $27$ rest-frame days before peak. This perhaps suggests that the UV/optical emission is not detected until the lines have formed, implying that the UV/optical emission in ASASSN-18pg may be the result of reprocessing of emission from the accretion disk. However, more TDEs with very early spectra, particularly of the TDE-Bowen class, are needed to see if there truly is a population of TDEs which show lines in all spectra, or whether TDEs simply exhibit lines at different timescales.

Our early observations also include two radio observations from ATCA obtained prior to peak light, roughly 2 weeks apart. The observations indicate the TDE was not detected in the radio, implying that if the TDE did launch a jet or outflow \citep[e.g.,][]{alexander16,velzen16}, it was not visible along our line-of-sight to the TDE.

Our late-time observations include both spectra and \swift{} observations obtained over 400 days after peak light. While the blackbody evolution of ASASSN-18pg at late times is consistent with those of other TDEs with similarly late observations, we do not detect X-ray emission in any epoch, nor do we detect any evolution in the X-rays as has been seen in other TDEs \citep[e.g.,][Hinkle et al. in prep.]{gezari17,holoien18a,velzen20}. Spectra obtained after 2019 March, roughly 9 months after peak light, show no evidence of features related to the TDE, nor is there any significant spectroscopic evolution after this time. 

Our dataset also includes the first published spectropolarimetric observations of a TDE, obtained roughly 10 rest-frame days prior to peak light. These observations find a polarization consistent with that of nearby stars and the line-of-sight Galactic extinction. If we interpret this as a $\sim 1\%$ upper limit on polarization from the source, this implies that the emission is relatively spherical. Based on the models for Type Ia SNe by \citet{hoflich91}, the axis ratio of the emission would have to be $\gtrsim 0.8$, or there is little scattering to produce the polarization. There are also no obvious changes in the polarization with wavelength, including any differences between the line and continuum emission. While spectropolarimetric observations are expensive, spectropolarimetry provides the only way of probing the symmetry of the emission, and so might open an important new window into TDE physics. Multiple epochs of spectropolarimetry are important both to look for changes in the symmetry and because changes in polarization can be measured without worrying about the contamination from foreground sources of polarization.

This dataset includes not only well-sampled observations after peak, as many UV/optical TDE datasets now do, but also the very early- and late-time data that has traditionally been missing in our observations of TDEs. These data are needed to differentiate between different emission models, and to test theoretical predictions for TDE emission. With surveys like ASAS-SN now finding TDEs earlier and more frequently, ASASSN-18pg should become one of many TDEs with similar datasets, hopefully resulting in a unified model of TDE emission.

\acknowledgments

We thank Y.-C. Pan for contributions to observing for this dataset.

We thank the Las Cumbres Observatory and its staff for its continuing support of the ASAS-SN project. ASAS-SN is supported by the Gordon and Betty Moore Foundation through grant GBMF5490 to the Ohio State University, and NSF grants AST-1515927 and AST-1908570. Development of ASAS-SN has been supported by NSF grant AST-0908816, the Mt. Cuba Astronomical Foundation, the Center for Cosmology and AstroParticle Physics at the Ohio State University,  the Chinese Academy of Sciences South America Center for Astronomy (CAS- SACA), the Villum Foundation, and George Skestos. 

KAA is supported by the Danish National Research Foundation (DNRF132). CSK and KZS are supported by NSF grants AST-1515927 and AST-181440. CSK, KZS and BJS are supported by NSF grant AST-1907570. MAT acknowledges support from the DOE CSGF through grant DE-SC0019323. BJS is also supported by NSF grants AST-1920392 and AST-1911074. Support for JLP is provided in part by FONDECYT through the grant 1151445 and by the Ministry of Economy, Development, and Tourism's Millennium Science Initiative through grant IC120009, awarded to The Millennium Institute of Astrophysics, MAS. TAT is supported in part by Scialog Scholar grant 24215 from the Research Corporation. PSC is grateful for support provided by NASA through the NASA Hubble Fellowship grant \#HST-HF2-51404.001-A awarded by the Space Telescope Science Institute, which is operated by the Association of Universities for Research in Astronomy, Inc., for NASA, under contract NAS 5-26555. KDF is supported by Hubble Fellowship grant HST-HF2-51391.001-A, provided by NASA through a grant from the Space Telescope Science Institute (STScI), which is operated by the Association of Universities for Research in Astronomy, Inc., under NASA contract NAS5-26555. DAHB is supported by the National Research Foundation (NRF) of South Africa. MG is supported by the Polish NCN MAESTRO grant 2014/14/A/ST9/00121. The UCSC transient team is supported in part by NSF grant AST-1518052, NASA/{\it Swift} grant 80NSSC19K1386, the Gordon \& Betty Moore Foundation, the Heising-Simons Foundation, and by a fellowship from the David and Lucile Packard Foundation to RJF.

Some of the observations were obtained using the Southern African Large Telescope (SALT) as part of the Large Science Programme on transients (2016-2-LSP-001; PI: Buckley). Polish participation in SALT is funded by grant no. MNiSW DIR/WK/2016/07.

Based on observations obtained at the international Gemini Observatory, a program of NSF’s OIR Lab, which is managed by the Association of Universities for Research in Astronomy (AURA) under a cooperative agreement with the National Science Foundation. on behalf of the Gemini Observatory partnership: the National Science Foundation (United States), National Research Council (Canada), Agencia Nacional de Investigaci\'{o}n y Desarrollo (Chile), Ministerio de Ciencia, Tecnolog\'{i}a e Innovaci\'{o}n (Argentina), Minist\'{e}rio da Ci\^{e}ncia, Tecnologia, Inova\c{c}\~{o}es e Comunica\c{c}\~{o}es (Brazil), and Korea Astronomy and Space Science Institute (Republic of Korea).

This research uses data obtained through the Telescope Access Program (TAP), which has been funded by the National Astronomical Observatories of China, the Chinese Academy of Sciences, and the Special Fund for Astronomy from the Ministry of Finance

This research was partially supported by the Australian Government through the Australian Research Council's Discovery Projects funding scheme (project DP200102471).

Parts of this research were supported by the Australian Research Council Centre of Excellence for All Sky Astrophysics in 3 Dimensions (ASTRO 3D), through project number CE170100013.

This research uses services or data provided by the NOAO Data Lab. NOAO is operated by the Association of Universities for Research in Astronomy (AURA), Inc. under a cooperative agreement with the National Science Foundation.

This research draws upon DECam data as distributed by the Science Data Archive at NOAO. NOAO is operated by the Association of Universities for Research in Astronomy (AURA) under a cooperative agreement with the National Science Foundation. 

This project used data obtained with the Dark Energy Camera (DECam), which was constructed by the Dark Energy Survey (DES) collaboration. Funding for the DES Projects has been provided by the U.S. Department of Energy, the U.S. National Science Foundation, the Ministry of Science and Education of Spain, the Science and Technology Facilities Council of the United Kingdom, the Higher Education Funding Council for England, the National Center for Supercomputing Applications at the University of Illinois at Urbana-Champaign, the Kavli Institute of Cosmological Physics at the University of Chicago, Center for Cosmology and Astro-Particle Physics at the Ohio State University, the Mitchell Institute for Fundamental Physics and Astronomy at Texas A\&M University, Financiadora de Estudos e Projetos,  Funda\c c\~{a}o Carlos Chagas Filho de Amparo, Financiadora de Estudos e Projetos, Funda\c c\~{a}o Carlos Chagas Filho de Amparo \`a Pesquisa do Estado do Rio de Janeiro, Conselho Nacional de Desenvolvimento Cient\'ifico e Tecnol\'ogico and the Minist\'erio da Ci\^encia, Tecnologia e Inova\c{c}\~ao, the Deutsche Forschungsgemeinschaft and the Collaborating Institutions in the Dark Energy Survey. 

The Collaborating Institutions are Argonne National Laboratory, the University of California at Santa Cruz, the University of Cambridge, Centro de Investigaciones En\'ergeticas, Medioambientales y Tecnol\'ogicas-Madrid, the University of Chicago, University College London, the DES-Brazil Consortium, the University of Edinburgh, the Eidgen\:ossische Technische Hochschule (ETH) Z\:urich, Fermi National Accelerator Laboratory, the University of Illinois at Urbana-Champaign, the Institut de Ci\`encies de l'Espai (IEEC/CSIC), the Institut de F\'isica d'Altes Energies, Lawrence Berkeley National Laboratory, the Ludwig-Maximilians Universit\:at M\:unchen and the associated Excellence Cluster Universe, the University of Michigan, the National Optical Astronomy Observatory, the University of Nottingham, the Ohio State University, the OzDES Membership Consortium, the University of Pennsylvania, the University of Portsmouth, SLAC National Accelerator Laboratory, Stanford University, the University of Sussex, and Texas A\&M University. 

Based on observations at Cerro Tololo Inter-American Observatory, National Optical Astronomy Observatory (Prop. ID 2018A-0251, PI D. Finkbeiner), which is operated by the Association of Universities for Research in Astronomy (AURA) under a cooperative agreement with the National Science Foundation.

\bibliography{bibliography.bib}
\bibliographystyle{apj}


\begin{deluxetable}{cccccc}[h!]
\tabletypesize{\footnotesize}
\tablecaption{Spectroscopic Observations of ASASSN-18pg}
\tablehead{
\colhead{Date} &
\colhead{Telescope} &
\colhead{Instrument} &
\colhead{Grating} &
\colhead{Slit} &
\colhead{Exposure Time}}
\startdata
2018 July 17.15 & SOAR 4.1-m & Goodman M1 & 400 l/mm & 1\farcs{00} & 1x900s \\
2018 July 18.93 & SALT 11.1-m & RSS & PG0300 & 1\farcs{50} & 1x1000s \\
2018 August 03.86 & SALT 11.1-m & RSS & PG0300 & 1\farcs{50} & 4x900s \\
2018 August 09.06 & du Pont 100-inch & WFCCD & Blue & 1\farcs{65} & 2x600s \\
2018 August 11.05 & du Pont 100-inch & WFCCD & Blue & 1\farcs{65} & 2x600s \\
2018 August 15.04 & du Pont 100-inch & WFCCD & Blue & 1\farcs{65} & 2x600s \\
2018 August 15.08 & Magellan Baade 6.5-m & IMACS f/2 & 300 l/mm & 0\farcs{90} & 3x300s \\
2018 September 13.97 & du Pont 100-inch & B\&C & 300 l/mm & 1\farcs{65} & 3x1000s \\
2018 September 28.00 & Gemini South 8.1-m & GMOS & R400 & 1\farcs{00} & 2x300s \\
2018 October 28.03 & SOAR 4.1-m & Goodman M1$+$M2 & 400 l/mm & 1\farcs{00} & 2x600s \\
2019 January 31.36 & du Pont 100-inch & WFCCD & Blue & 1\farcs{65} & 3x600s \\
2019 February 12.36 & SOAR 4.1-m & Goodman M1 & 400 l/mm & 1\farcs{00} & 1x1200s \\
2019 March 04.33 & du Pont 100-inch & WFCCD & Blue & 1\farcs{65} & 3x900s \\
2019 March 13.32 & SOAR 4.1-m & Goodman M1$+$M2 & 400 l/mm & 1\farcs{00} & 2x1500s \\
2019 March 28.32 & SOAR 4.1-m & Goodman M1$+$M2 & 400 l/mm & 1\farcs{00} & 2x1800s \\
2019 March 30.30 & Magellan Clay 6.5-m & LDSS-3 & VPH-All & 1\farcs{00} blue & 4x600s \\
2019 May 11.23 & du Pont 100-inch & WFCCD & Blue & 1\farcs{65} & 4x1800s \\
2019 May 11.29 & SOAR 4.1-m & Goodman M2 & 400 l/mm & 1\farcs{00} & 1x1800s \\
2019 June 04.14 & du Pont 100-inch & WFCCD & Blue & 1\farcs{65} & 3x900s \\
2019 June 06.99 & SOAR 4.1-m & Goodman M1$+$M2 & 400 l/mm & 1\farcs{00} & 2x1050s \\
2019 August 07.04 & SOAR 4.1-m & Goodman M1 & 400 l/mm & 1\farcs{00} & 1x1800s \\
2019 September 24.99 & du Pont 100-inch & WFCCD & Blue & 1\farcs{65} & 3x1200s \\
\enddata 
\tablecomments{Date, telescope, instrument, grating, slit size, and exposure time for each of the spectroscopic observations obtained of ASASSN-18pg for the initial classification of the transient and as part of our follow-up campaign.} 
\label{tab:spec_details} 
\end{deluxetable}

\end{document}